\documentclass[useAMS,usenatbib]{mn2e}

\usepackage{graphicx,rotating,lscape,color}

\title[Stars with TiO bands in emission]
{Magellanic Cloud stars with TiO bands in emission: \\binary post-RGB/AGB stars or young stellar objects?}

\author[P. R. Wood, D. Kamath and H. Van Winckel]{P. R. Wood$^{1}$\thanks{E-mail:
wood@mso.anu.edu.au (PRW); devika13@gmail.com (DK); Hans.VanWinckel@ster.kuleuven.be (HVW)},
D. Kamath$^{1}$ and H. Van Winckel$^{2}$\\
$^{1}$Research School of Astronomy and Astrophysics, Australian National University,
Cotter Road, Weston Creek ACT 2611, Australia\\
$^{2}$Instituut voor Sterrenkunde, K.U. Leuven, Celestijnenlaan 200D, 3001 Leuven, Belgium
}
\begin{document}

\date{}

\pagerange{\pageref{firstpage}--\pageref{lastpage}} \pubyear{2002}

\maketitle

\label{firstpage}

\begin{abstract}

Fourteen stars from a sample of Magellanic Cloud objects selected to
have a mid-infrared flux excess have been found to also show TiO bands in
emission.  The mid-infrared dust emission and the TiO band emission
indicate that these stars have large amounts of hot circumstellar dust and
gas in close proximity to the central star.  The luminosities of the
sources are typically several thousand L$_{\odot}$ while the effective
temperatures are $\sim$4000-8000 K which puts them blueward of the
giant branch.  Such stars could be post-AGB stars of mass
$\sim$0.4--0.8\,M$_{\odot}$ or pre-main-sequence stars 
(young stellar objects) with masses of
$\sim$7-19\,M$_{\odot}$.  If the stars are pre-main-sequence stars,
they are substantially cooler and younger than stars at the birth line
where Galactic protostars are first supposed to become optically visible out of
their molecular clouds.  They should therefore be hidden
in their present evolutionary state, although this problem may be
overcome if asymmetries are invoked or if the reduced metallicity 
of the SMC and LMC compared to the Galaxy makes the circumstellar material 
more transparent.  The second explanation for these
stars is that they are post-AGB or post-RGB stars that have recently
undergone a binary interaction when the red giant of the binary system
filled its Roche lobe.  Being oxygen-rich, they have gone through this
process before becoming carbon stars.  Most of the stars vary slowly
on timescales of 1000 days or more suggesting a changing circumstellar
environment. Apart from the slow variations, most stars also show variability
with periods of tens to hundreds of days.  One star shows a period that is
rapidly decreasing and we speculate that this star may have accreted a
large blob of gas and dust onto a disk whose orbital radius is shrinking rapidly.
Another star has Cepheid-like pulsations of rapidly increasing amplitude
suggesting a rapid rate of evolution.  Seven stars show quasi-periodic 
variability and one star has a light curve similar to that of an
eclipsing binary.
\end{abstract}



\section{Introduction}

As part of an optical spectral survey of post-AGB candidates in the Magellanic
Clouds, we have discovered 14 stars that show bandheads of the TiO molecule
in emission.  The post-AGB candidates were selected because they had excess
mid-infrared emission in their spectral energy distributions (SEDs), indicating the
presence of circumstellar dust, and presumably gas.

Normal red giant or main-sequence stars with effective temperatures
T$_{\rm eff} \la 3800$\,K (the M stars) show TiO bands in absorption.
However, a small number of stars have previously been found to show TiO bands
in emission.  \citet{cov11} and \citet{hil12} have found 3 nearby
young stellar objects (YSOs) which show TiO bands 
in emission.  These stars have various
other emission lines, especially H$\alpha$, and \citet{cov11} and
\citet{hil12} argue that their YSOs have accretion disks and that
the TiO band emission comes from dense circumstellar gas with n $\ga$
10$^{10}$ cm$^{-3}$ and $T \sim 1400-4000\,K$ in the accretion disk.
An earlier study by \citet{zic89} found 4 Be stars in which they
tentatively identified emission in the TiO bandhead at 6159\,\AA\,
\citep[our spectra of some of these Be stars, to be published elsewhere, 
show additional bandheads of TiO
at longer wavelengths, confirming the findings of][]{zic89}.  The
Be stars also show a mid-IR flux excess in their SEDs and are known to
be surrounded by disks of dust and gas \citep[e.g.][]{por03}.
\citet{zic89} estimate a gas density greater than $\sim$10$^9$
cm$^{-3}$ for the circumstellar gas in their stars with TiO bands in
emission.  Both YSOs and Be stars which show TiO band
emission have been found to also show the first overtone band of the
CO molecule in emission at 2.3\,$\mu$m \citep{zic89,hil12}.

The common feature linking the YSOs and Be stars is a
circumstellar disk and it therefore seems that a circumstellar disk of
gas and dust is an essential component for the production of TiO band
emission and the unusual temperature structure it requires.  A common feature
of binary post-AGB stars is the presence of a circumbinary disk
\citep[e.g.][]{van04,der06}, so these stars could also potentially show TiO
bands in emission.  Here, we describe objects with TiO band emission in
the Magellanic Clouds that are possible post-AGB or post-RGB star binaries, or
YSOs.

\section{Observations and data reduction}

Full details of the selection of objects, spectral observations, data
reduction and estimation of luminosity and $T_{\rm eff}$ are given by
\citet{kam13}.  In brief, post-AGB candidates were selected using
photometry of Magellanic Cloud stars from the Spitzer Space Telescope
surveys SAGE \citep{mei06} and SAGE-SMC \citep{gor11} combined with
optical UBVI photometry from \citet{zar02} for the SMC and
\citet{zar04} for the LMC.  Candidates were selected mostly for their
strong 24 $\mu$m and/or 8 $\mu$m flux excesses.  Optical spectra were
taken with the multi-fibre AAOmega spectrograph \citep{smi04} and have
a resolution of $\sim$1300 and a wavelength range of $\sim$3700--8800\,\AA.  
A computer program was created which automatically derived
$T_{\rm eff}$, $\log g$ and [Fe/H] by comparing the observed spectra
to synthetic spectra from \citet{mun05}.  Given the large amount
of molecular band emission superimposed on our observed spectra,
additional uncertainties are associated with the parameters derived by
this procedure.  We therefore also made eye-estimates of the spectral type
using the spectral features in the interval $\sim$3700-4700\,\AA~ 
shown in \citet{gc09} (predominantly Balmer lines, 
Ca II H and K lines and the G-band).
$T_{\rm eff}$ was then computed using the ($T_{\rm eff}$, spectral
type) relation given by \citet{pic98}\footnote{A convenient
tabulation is given at \\
http://www.stsci.edu/hst/HST\_overview/documents/synphot/ AppA\_Catalogs5.html}.
Table~\ref{tab1} lists these values of $T_{\rm eff}$ along with spectral types 
and the automatically derived values of $T_{\rm eff}$, $\log g$ and [Fe/H].  
In general, the two values of $T_{\rm eff}$ are reasonably similar.

The luminosities of the central stars were computed in two ways.
First, after removing a foreground extinction corresponding to
E(B-V)=0.08 and 0.12 \citep{kel06} for the LMC and SMC, respectively,
and using the extinction law of \citet{car89}, the apparent luminosity
was computed by integrating under the SED made from the photometry described
above, along with WISE photometry in the W1--4 bands \citep{wri10}.
The absolute luminosities $L_{\rm obs}$ were then obtained by applying
distance modulii of 18.54 and 18.93 \citep{kel06} for the LMC and
SMC, respectively.  We also note that the heliocentric radial
velocities of all the stars (Table~\ref{tab1}) are consistent with membership of the
Magellanic Clouds so the adopted distance modulii are appropriate.

For circumstellar dust that is not in a spherically symmetric
distribution, $L_{\rm obs}$ could be either an overestimate
or an underestimate.  For example, in the case of a dense disk obscuring
the central star but only capturing a fraction of the 4$\pi$
steradians of photospheric emission, $L_{\rm obs}$ will be an
underestimate of the total emission while if the disk is oriented so
that its pole points to the observer, $L_{\rm obs}$ will be an
overestimate.

The second method of computing the luminosity of the central star has several
steps.  Firstly, the intrinsic colours of the central star were
derived from the estimated $T_{\rm eff}$ (the automatically derived
$T_{\rm eff}$ was used).  The reddening was then derived by finding
the value of E(B-V) that minimized the sum of the squared differences
between the dereddened observed and the intrinsic B, V, I and J
magnitudes.  The \citet{car89} extinction law\footnote {It is possible
that the circumstellar extinction law is different from the
interstellar extinction law but we have not explored this
possibility.  For example, \citet{coh04} adopt a gray circumstellar
extinction law for the Red Rectangle due to very large grains in the
circumbinary disk.} was used.  Using the derived E(B-V), the
observed magnitudes were corrected for extinction.  Then the BVIJ
fluxes of the best-fit model atmosphere from the $T_{\rm eff}$
estimation procedure were normalized to the corrected BVIJ fluxes.
The bolometric correction to V for the model atmosphere, coupled with
the distance modulii to the LMC and SMC then allowed the derivation of
the photospheric luminosity $L_{\rm phot}$.  We note that this
luminosity should be free from errors caused by asymmetry in the
circumstellar dust distribution except when there is a substantial
non-photospheric flux in the BVIJ bands from emission or scattering by
circumstellar matter.  This is known to occur in some cases e.g. in
the bipolar post-AGB star known as the Red Rectangle \citep{coh04}.
There will also be uncertainties in the luminosity arising from errors
in the estimation of $T_{\rm eff}$.  A comparison of $L_{\rm phot}$
and $L_{\rm obs}$ gives some estimate of the errors involved: 
a large difference between the two values, which occurs in two cases,  
suggests a non-spherical 
distribution of circumstellar dust.  Both
$L_{\rm phot}$ and $L_{\rm obs}$ are listed in Table~\ref{tab1}
together with the total reddening E(B-V) to the photosphere estimated
as described above.


\begin{landscape}
\begin{table}
\caption{Properties of the objects}
\medskip
\begin{tabular}{lllcrclcrrclcc}
\hline
   Name              & SpT &$T_{\rm eff}$& $v$& $T_{\rm eff}$ & $\log g$ & [Fe/H] & E(B-V) & $L_{\rm obs}$& $L_{\rm phot}$& Star  & Period                 & H$\alpha$& Li\\
                     &     & SpT  &km s$^{-1}$&               &  cgs     &        &        &  L$_{\odot}$ &  L$_{\odot}$  & Type  & ~days                  & emission & absorption\\
\hline                                                                                                                     
J004805.01-732543.0  & K   & 4-5000    & 199  &   3849        &     0.0  &  -1.05 &   0.16 &    3492      &    1960       & p-RGB & 1150$\rightarrow$400  &yes       &?\\
J004843.76-735516.8  & G0  & 5500      & 183  &   5337        &     0.8  &  -1.46 &   0.09 &   10067      &    7677       & ?     & 114,164               &no        &no\\
J005355.00-731900.9  & K2  & 4250      & 189  &   4140        &     0.0  &  -1.38 &   0.30 &    2869      &    2385       & p-RGB & 193                   &yes       &strong\\
J005504.57-723451.1  & A2  & 8900      & 183  &   7480        &     3.0  &  -1.0  &   0.10 &    2148      &    2220       & PMS & -                       & resolved &no\\
J005514.24-732505.3  & F5  & 6640      & 163  &   6404        &     2.5  &  -1.0  &   0.18 &    3486      &    743        & PMS & 206                     &yes       &no\\
J005529.48-715312.2  & F0  & 7690      & 178  &   6457        &     2.1  &  -0.39 &   0.37 &    5141      &    6668       & PMS & -                       &no        &no\\
J010324.36-723803.5  & G0  & 5500      & 189  &   4447        &     0.0  &  -1.16 &   0.06 &   28384      &    21757      & p-AGB & -                     &no        &no\\
J010628.81-715204.8  & A5  & 8450      & 191  &   7443        &     2.5  &  -1.0  &   0.24 &    5031      &    4988       & PMS & 33.5                    &P Cyg     &no\\
J010929.79-724820.6  & K5  & 4000      & 129  &   4173        &     0.7  &  -1.02 &   0.33 &    3079      &    3732       & ?   & 444$^a$,100,69.7        &no        &?\\
J050747.45-684351.2  & K0  & 4850      & 267  &   5430        &     0.5  &  -1.36 &   0.34 &    1304      &    2068       & p-RGB & -                     &P Cyg     &?\\
J051155.66-693020.6  & K0  & 4850      & 245  &   4096        &     0.0  &  -0.9  &   0.04 &    1865      &    1618       & p-RGB & -                     &no        &?\\
J051516.28-685539.7  & K:  & 4-5000    & 311  &   3878        &     0.0  &  -1.06 &   0.19 &    3529      &    2943       & p-AGB & 380$^a$,124           &no        &strong\\
J052023.97-695423.2  & GK::& 4-6000    & 324  &   5244        &     2.0  &  -2.5  &   0.17 &    2907      &    626        & PMS   & -                     &resolved  &no\\
J052230.40-685923.9  & GK::& 4-6000    & 299  &   4493        &     1.6  &  -0.97 &   0.51 &    2876      &    3375       & PMS   & 46                    &no        &strong\\
\hline
\end{tabular}
\begin{flushleft}
Notes: ``SpT'' is the eye-estimated spectral type, ``$v$'' is the heliocentric radial velocity and ``Star Type''
is the estimated evolutionary status based on $\log g$ \\(p-RGB for a binary post-RGB star, 
p-AGB for a binary post-AGB star, PMS for a pre-main-sequence star and ? if we could
not derive the \\type from $\log g$).  ``Star Type'' is quite uncertain (see text).\\
$^a$ Long secondary period.
\end{flushleft}
\label{tab1}
\normalsize
\end{table}
\end{landscape}

\begin{landscape}
\begin{table}
\caption{Photometry of the objects}
\medskip
\begin{tabular}{lcccccccccccccccc}
\hline
Name                &    U   &    B   &    V   &    I   &    J   &    H   &    K   &   W1   &[3.6]   &[4.5]   &   W2   &[5.8]   &[8.0]   &   W3   &   W4   & [24]\\ 
\hline                                                                                                                                                  
J004805.01-732543.0 &  ...   & 19.321 & 17.282 & 15.838 & 13.853 & 12.737 & 12.121 & 11.055 & 10.832 & 10.278 & 10.315 &  9.733 &  8.968 &  8.287 &  7.080 &  7.275 \\
J004843.76-735516.8 & 15.714 & 15.091 & 14.434 & 13.290 & 12.733 & 12.023 & 11.572 & 10.733 & 10.670 & 10.289 & 10.219 &  9.948 &  9.442 &  8.790 &  8.117 &  7.942 \\
J005355.00-731900.9 & 19.640 & 18.682 & 16.955 & 14.982 & 13.909 & 12.928 & 12.454 & 11.611 & 11.437 & 10.988 & 10.964 & 10.456 &  9.902 &  9.286 &  7.360 &  8.451 \\
J005504.57-723451.1 & 15.607 & 15.824 & 15.592 & 15.217 & 14.876 & 14.557 & 14.211 & 13.434 & 13.184 & 12.721 & 12.824 & 12.341 & 11.672 & 11.313 &  9.379 &  ...   \\
J005514.24-732505.3 & 17.840 & 17.548 & 17.072 & 16.470 & 15.776 & 14.205 & 12.699 & 10.702 & 10.347 &  9.661 &  9.652 &  8.999 &  8.338 &  7.794 &  6.791 &  6.716 \\
J005529.48-715312.2 & 16.375 & 16.029 & 15.239 & 14.310 & 13.683 & 13.191 & 12.739 & 11.370 & 11.314 & 10.586 & 10.535 &  9.835 &  8.775 &  7.787 &  5.938 &  5.855 \\
J010324.36-723803.5 & 15.733 & 14.904 & 13.556 & 12.168 & 11.373 & 10.661 & 10.374 &  9.567 &  9.510 &  9.018 &  8.971 &  8.423 &  7.831 &  7.272 &  6.377 &  6.408 \\
J010628.81-715204.8 & 15.719 & 15.411 & 15.135 & 14.570 & 14.092 & 13.680 & 12.997 & 11.455 & 11.178 & 10.469 & 10.429 &  9.767 &  8.853 &  7.878 &  5.694 &  5.626 \\
J010929.79-724820.6 & 20.114 & 18.150 & 16.469 & 14.641 & 13.512 & 12.602 & 12.398 & 12.221 & 12.186 & 11.955 & 12.110 & 11.612 & 11.222 & 11.018 &  9.398 &  9.657 \\
J050747.45-684351.2 & 17.751 & 17.272 & 16.172 & 15.099 & 14.267 & 13.640 & 13.408 & 12.333 & 12.620 & 12.193 & 12.044 & 11.756 & 11.184 & 11.364 &  9.856 &  9.542 \\
J051155.66-693020.6 & 19.541 & 17.898 & 16.114 & 14.843 & 13.946 & 13.254 & 12.821 & 12.025 & 11.730 & 11.097 & 11.245 & 10.539 &  9.794 &  9.167 &  8.869 &  7.974 \\
J051516.28-685539.7 & 20.124 & 18.046 & 16.190 & 14.369 & 13.236 & 12.306 & 12.011 & 11.099 & 11.013 & 10.358 & 10.339 &  9.687 &  8.838 &  8.066 &  6.813 &  6.854 \\
J052023.97-695423.2 & 18.424 & 18.246 & 17.063 & 16.090 & 15.630 & 14.937 & 13.803 & 11.723 &  ...   & 10.341 & 10.347 &  9.491 &  8.442 &  7.027 &  3.786 &  3.620 \\
J052230.40-685923.9 &  ...   & 18.147 & 16.547 & 14.584 & 13.493 & 12.510 & 12.030 &  ...   & 10.927 & 10.492 &  ...   & 10.009 &  9.468 &  ...   &  ...   &  7.933 \\
\hline
\end{tabular}
\begin{flushleft}
Notes: U, B, V and I magnitudes from \citet{zar02}, J, H, K, [3.6], [4.5], [5.8], [8.0] and [24] from the SAGE \citep{mei06} and SAGE-SMC \citep{gor11} \\
catalogs and W1, W2, W3 and W4 from the WISE catalog.  Where no magnitude exists, a value of  ...   is given.
\end{flushleft}
\label{mags_tab}
\normalsize
\end{table}
\end{landscape}


\begin{figure*}
\centering
\begin{minipage}[]{13cm}
\includegraphics[width=6.5cm]{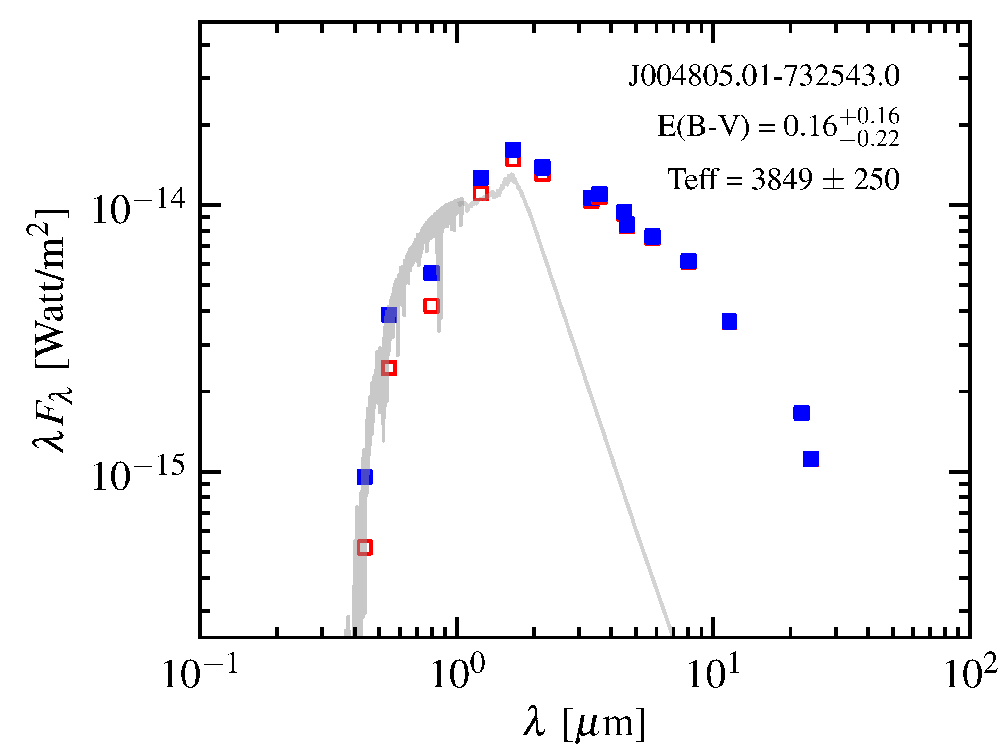}
\vspace{3mm}
\includegraphics[width=6.5cm]{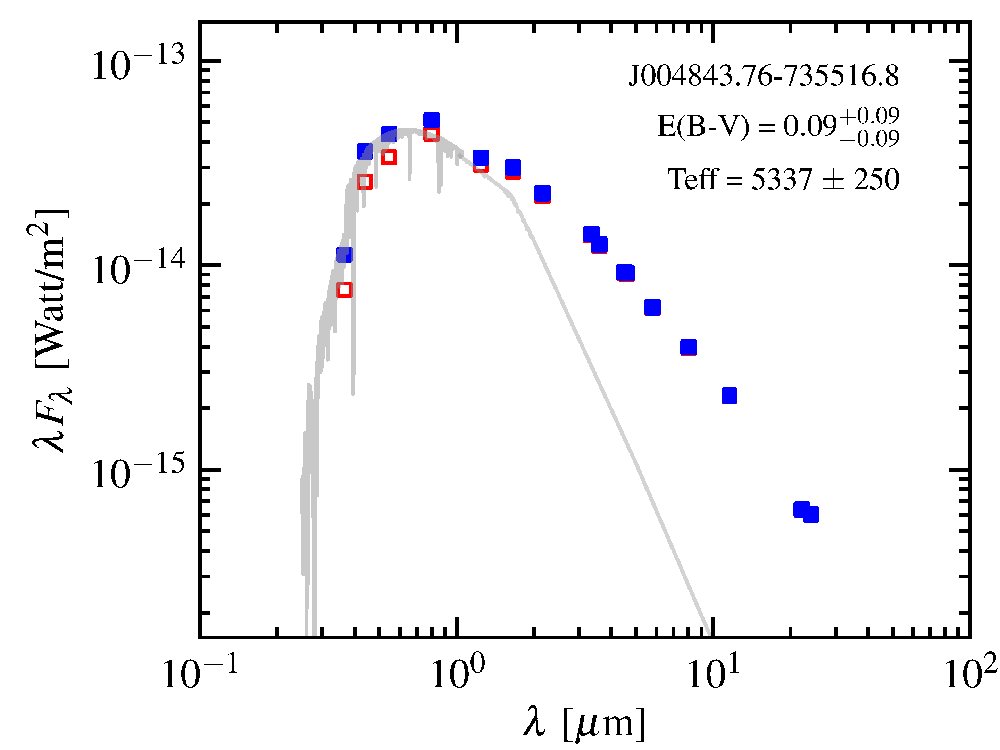}
\includegraphics[width=6.5cm]{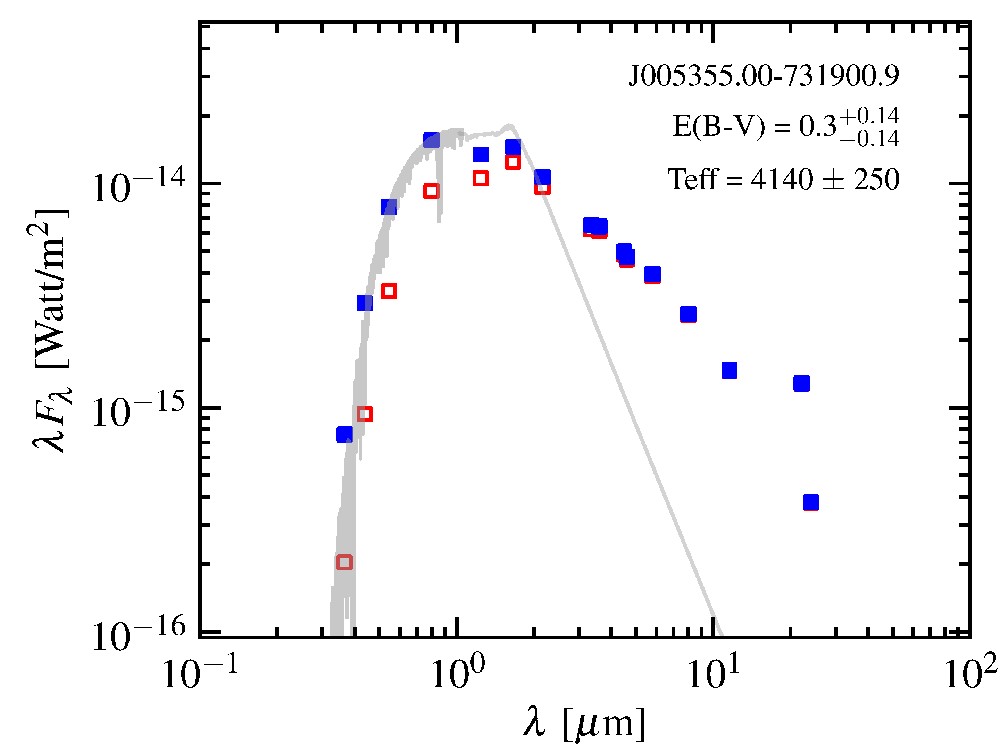}
\vspace{3mm}
\includegraphics[width=6.5cm]{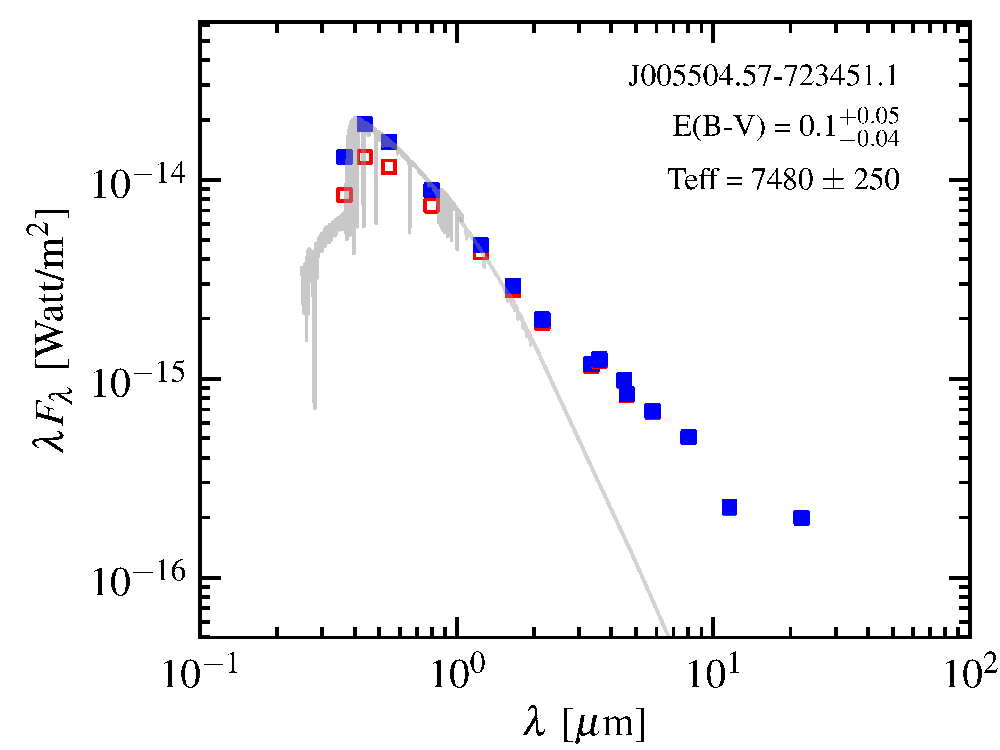}
\includegraphics[width=6.5cm]{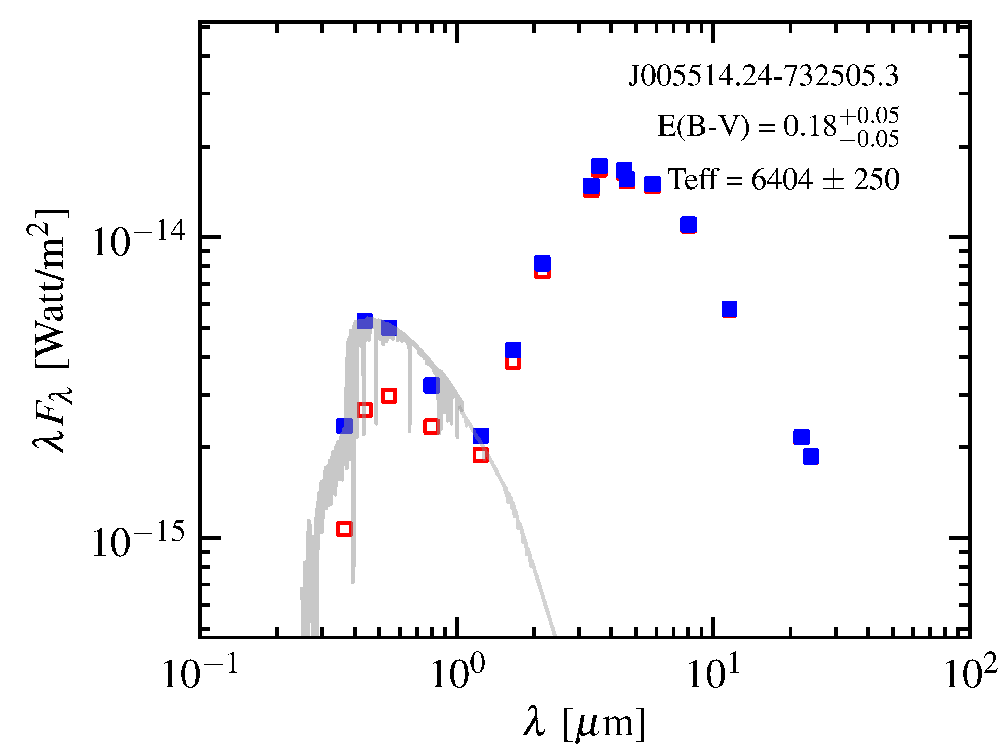}
\vspace{3mm}
\includegraphics[width=6.5cm]{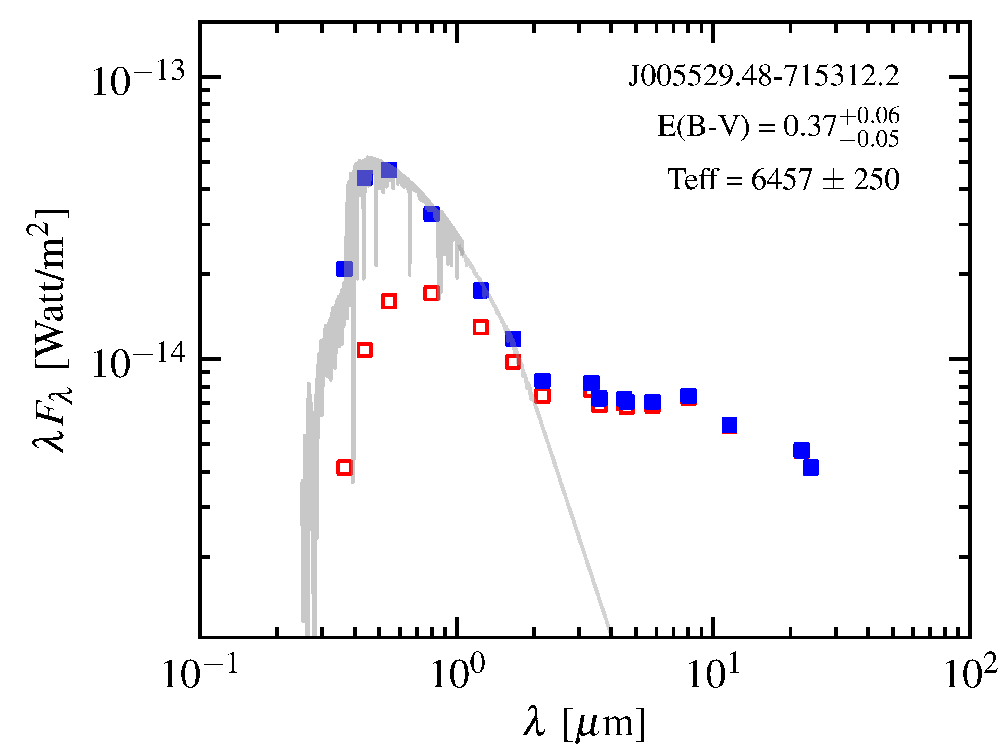}
\end{minipage}
\caption{SEDs of the first 6 sources. The red open squares show the observed broadband 
photometry while the blue filled squares show the dereddened  
photometry. For wavelengths up to 1.05$\mu$m, the 
best-fit Munari synthetic spectrum is plotted while for longer wavelengths
the low resolution flux distribution from the corresponding atmospheric model 
of \citet{cas04} is plotted.}
\label{seds_fig1}
\end{figure*}

\begin{figure*}
\centering
\begin{minipage}[]{13cm}
\includegraphics[width=6.5cm]{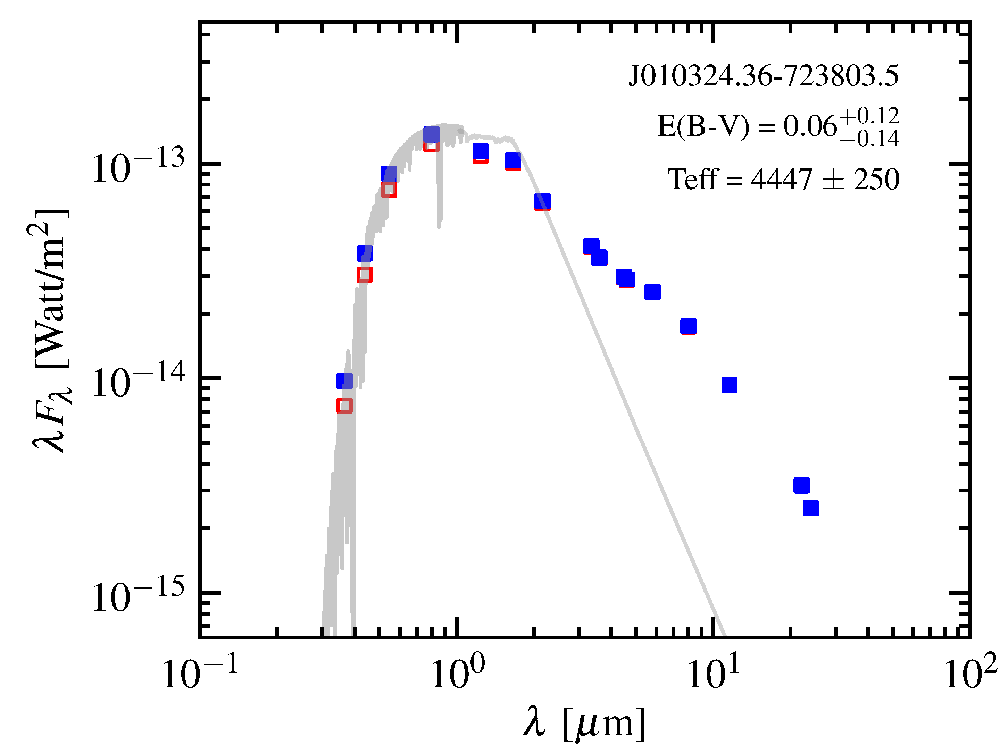}
\vspace{3mm}
\includegraphics[width=6.5cm]{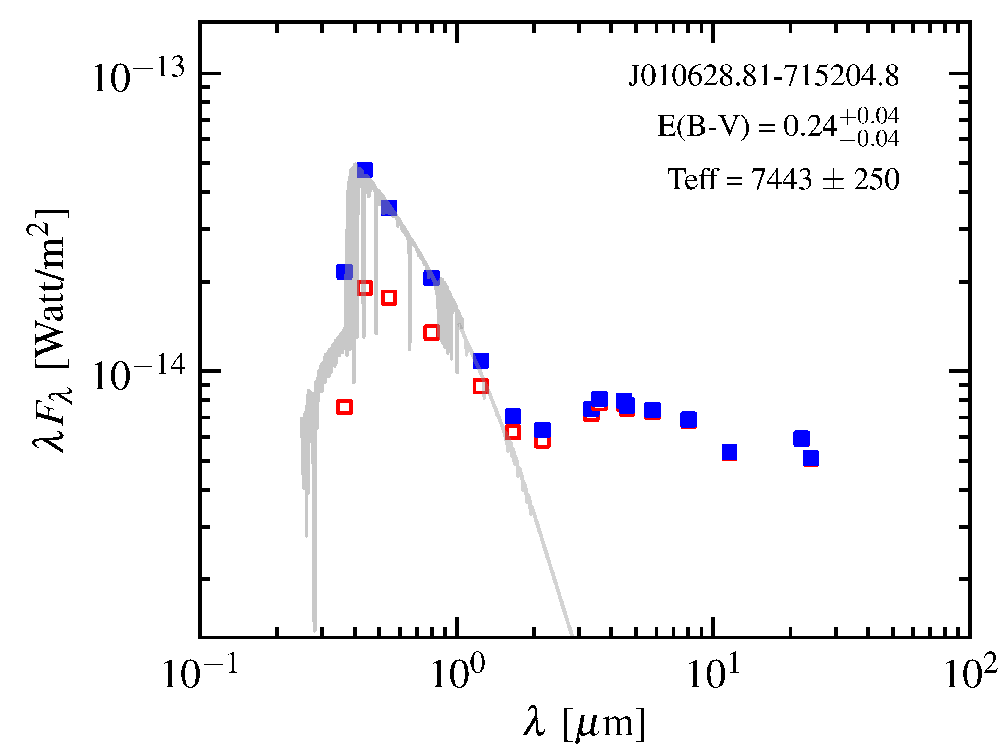}
\includegraphics[width=6.5cm]{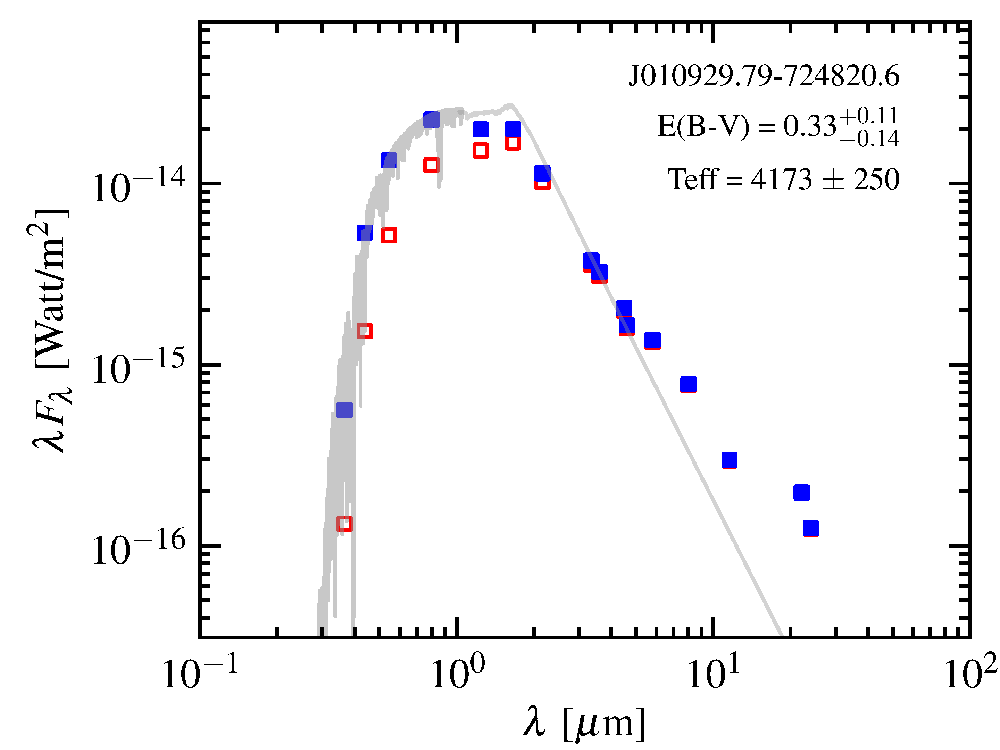}
\vspace{3mm}
\includegraphics[width=6.5cm]{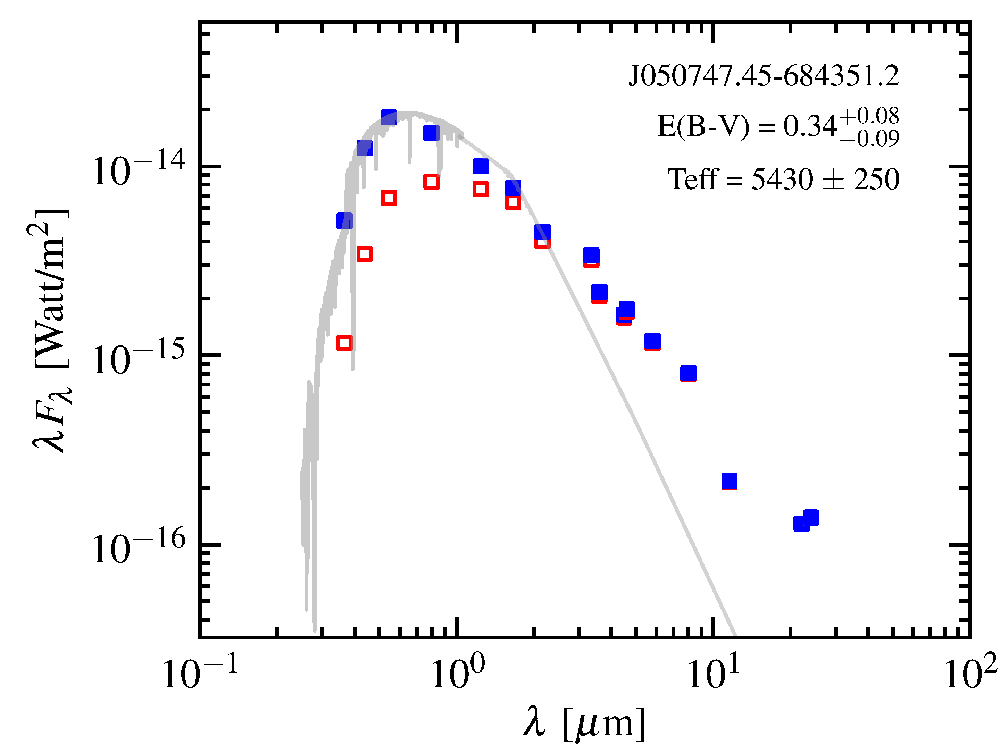}
\includegraphics[width=6.5cm]{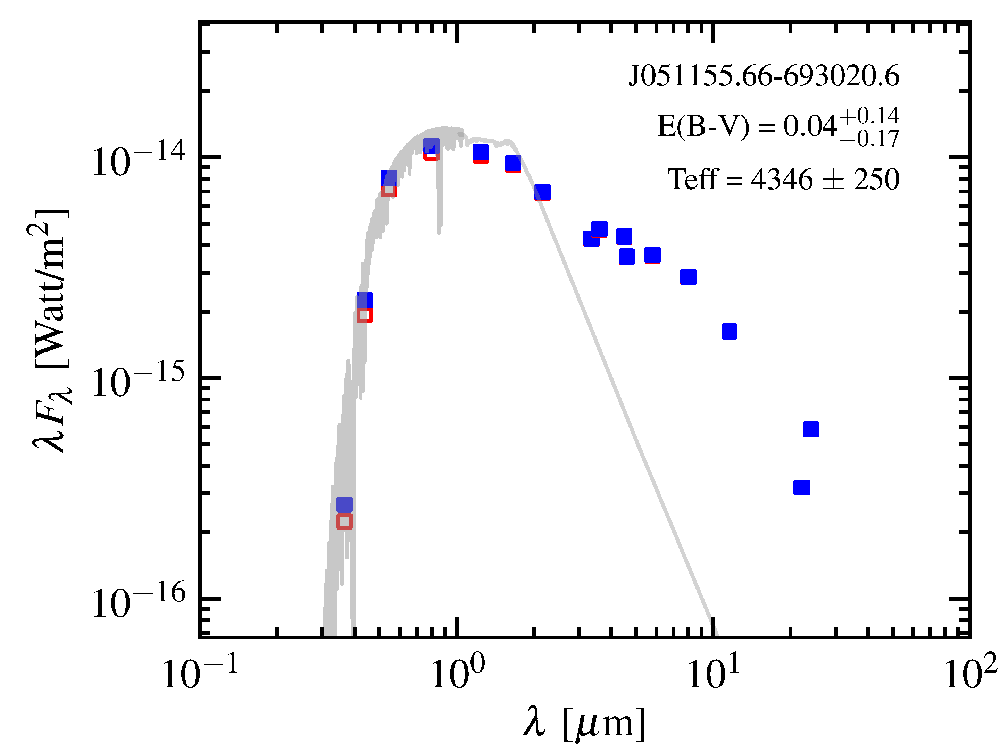}
\vspace{3mm}
\includegraphics[width=6.5cm]{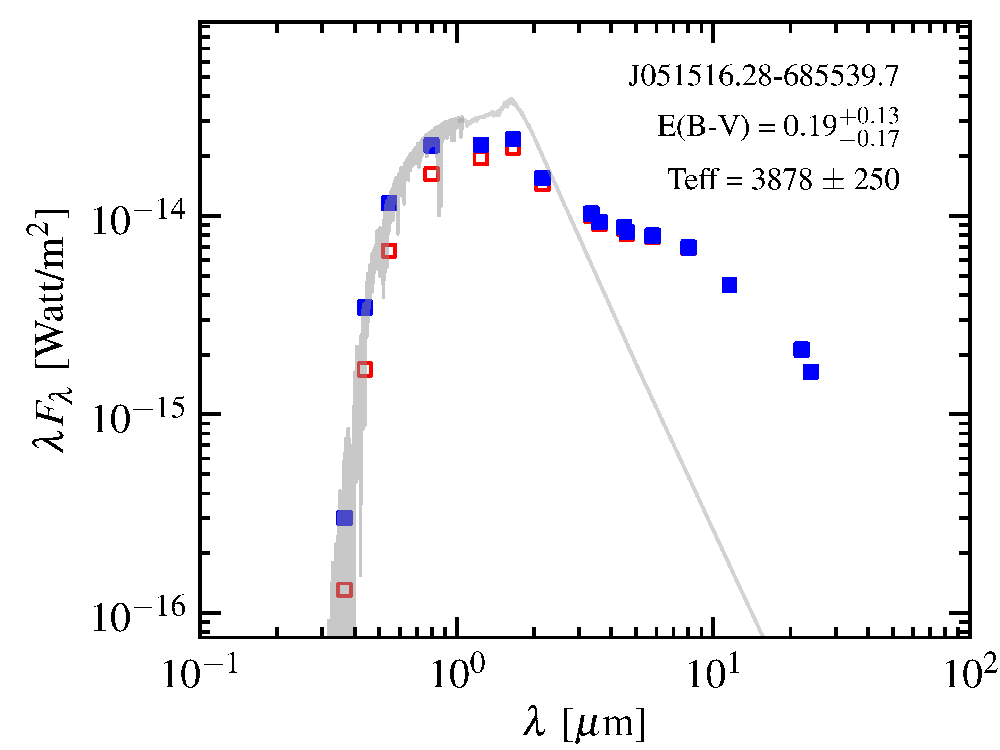}
\includegraphics[width=6.5cm]{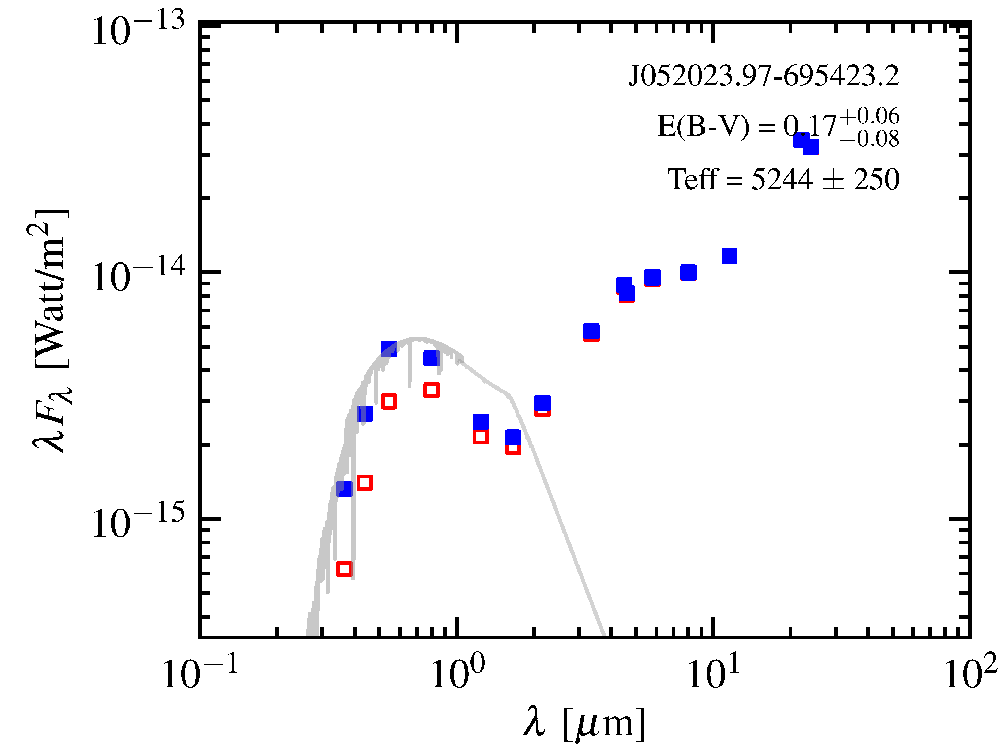}
\includegraphics[width=6.5cm]{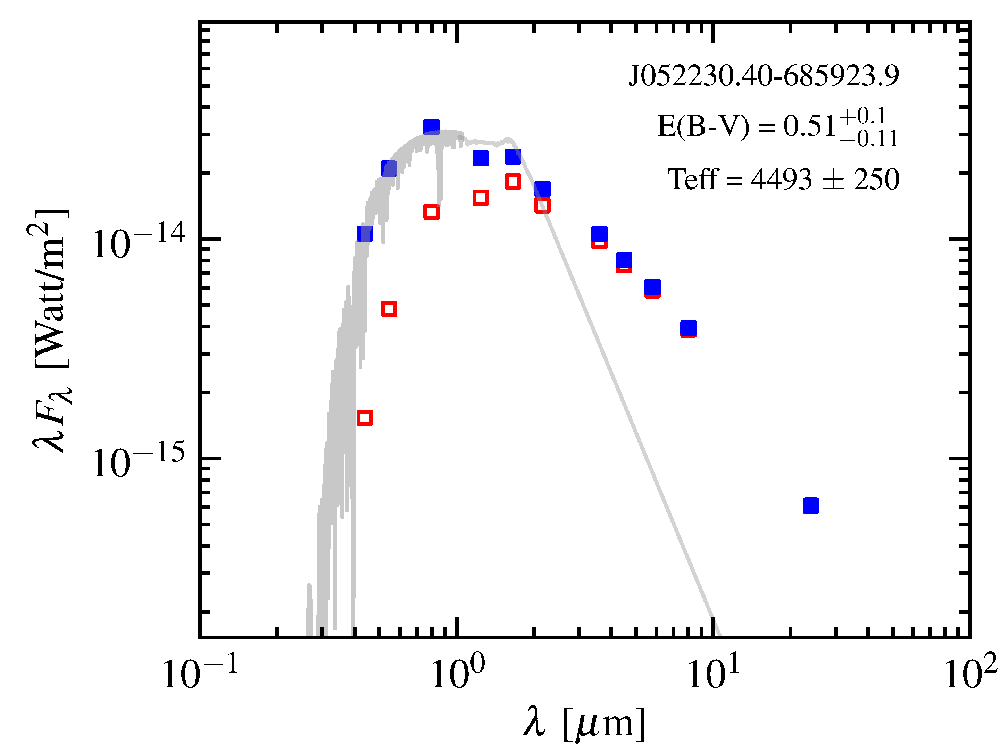}
\end{minipage}
\caption{SEDs of the remaining 8 sources.}
\label{seds_fig2}
\end{figure*}

\section{Results}

\subsection{Spectral energy distributions}\label{sed_sec}

The observed broadband magnitudes of the objects from U to 24\,$\mu$m
are given in Table~\ref{mags_tab}.  In Figures~\ref{seds_fig1} and
\ref{seds_fig2}, the SEDs corresponding to the observed magnitudes are
shown by red points while the SEDS corresponding to the dereddened
magnitudes are shown by blue points.  Also shown in each plot is the
energy distribution of the best-fitting model atmosphere.  In most
cases, the dereddened SED consists of the photospheric emission from
the central star and moderate amounts of excess mid-IR emission at
wavelengths longer than $\sim$1.2\,$\mu$m.  This is as expected for
stars surrounded by dust that absorbs a fraction of the photospheric
flux and re-emits it in the near- to mid-IR.  The dust could be in a
disk or a close-in circumstellar shell.

The SED of the star J005514.24-732505.3 is very unusual, with the
mid-IR luminosity dominating the photospheric luminosity.  The
luminosity $L_{\rm obs}$ obtained by integrating under the SED for
J005514.24-732505.3 is $\sim$5 times the luminosity $L_{\rm phot}$
estimated for the photosphere of the optically visible star.  For a
single star, this luminosity ratio requires a special geometry with a
thick disk seen edge-on.  The disk obscures the central star whose
optical light is seen mainly by scattering of light emerging through
the poles of the disk.  From detailed modelling, \citet{men02} state
that $L_{\rm obs}$ can be several times $L_{\rm phot}$ in this case.  Another
possibility is that there is a second {\it luminous} star embedded with the observed
star, or there is another independent object coincident on the sky
with the optically observed star.  The object J052023.97-695423.2 has
a similarly large $L_{\rm obs}$.  For it, $L_{\rm obs}$ is $\sim$2.8 times the
luminosity $L_{\rm phot}$.  J052023.97-695423.2 has previously 
been classified as a high-probability YSO candidate - see Section~\ref{hrd_sect}.

\subsection{Optical spectra}

\begin{figure*}
\includegraphics[scale=0.80]{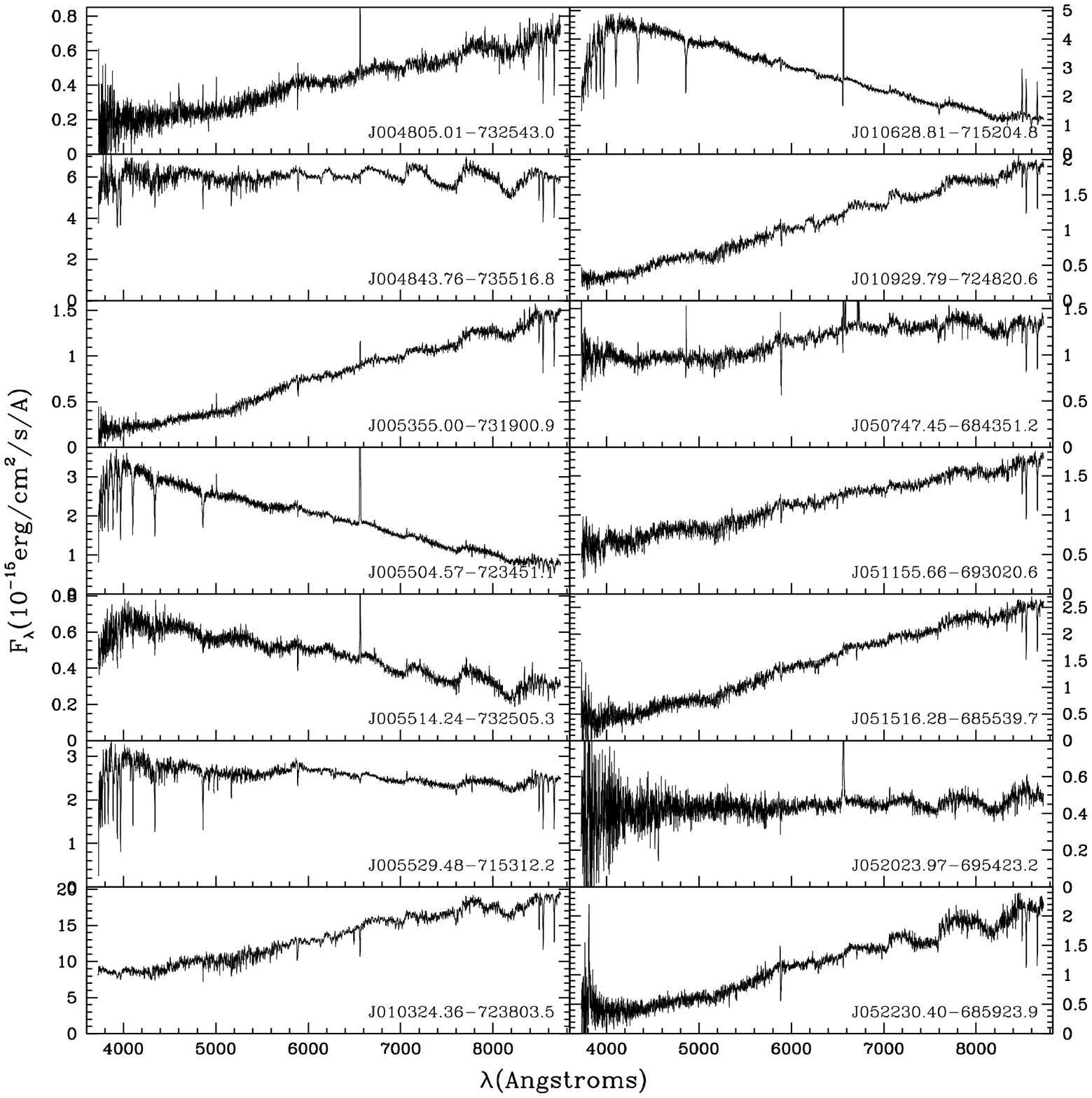}
\caption{The spectra of the objects showing TiO bands in emission.}
\label{full_spec}
\end{figure*}

\begin{figure*}
\includegraphics[scale=0.80]{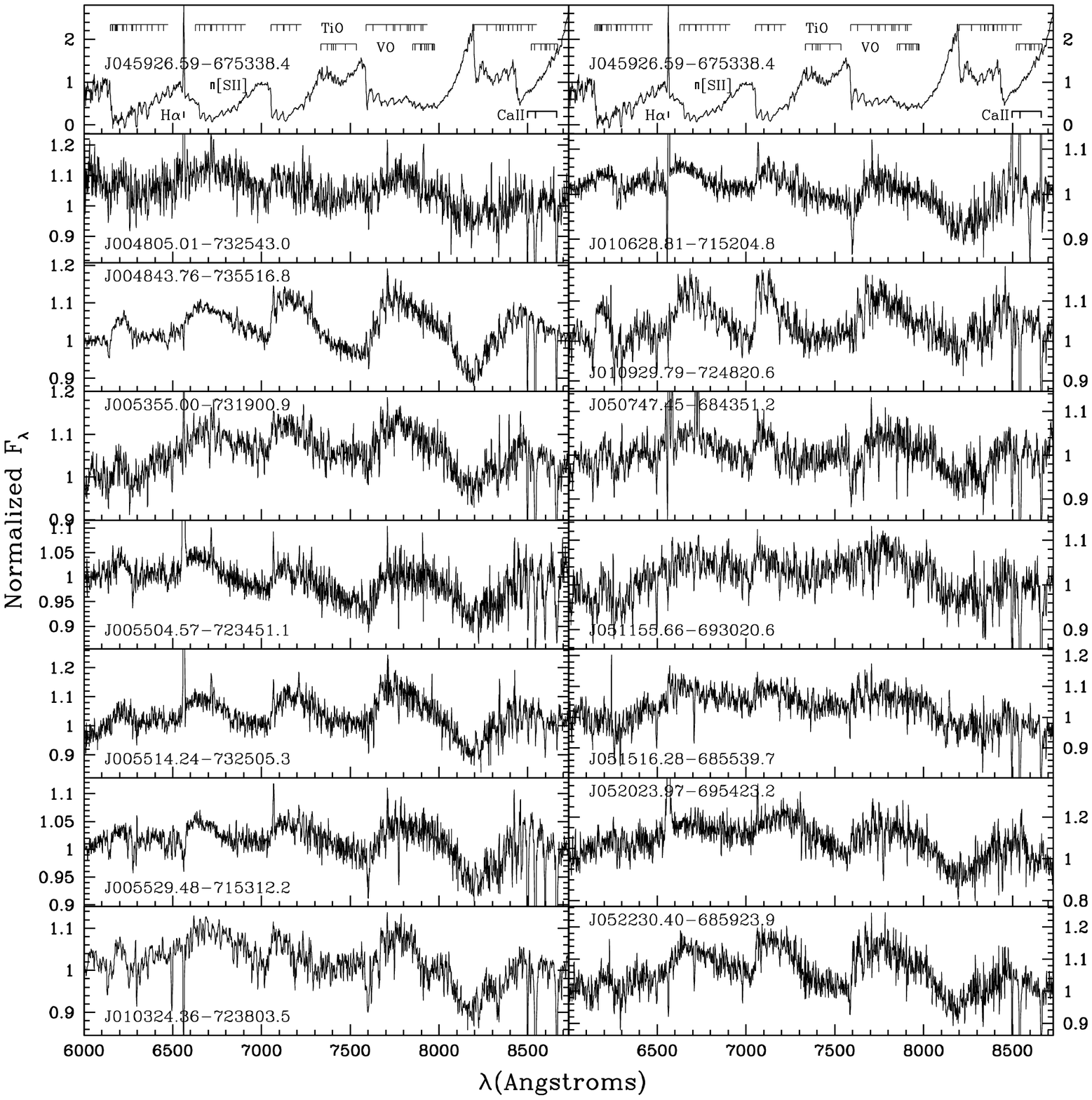}
\caption{The continuum-normalized spectra from 6000--8700\,\AA~.
Also shown is the spectrum of a late M star with the positions of band heads of
the TiO and VO molecules marked   The positions of H$\alpha$, the [SII] lines at 6717
and 6731\,\AA~ and the Ca triplet lines are also shown.}
\label{tio_spec}
\end{figure*}

The full observed optical spectra for the objects studied here are
shown in Figure~\ref{full_spec}.  The absolute flux values for the
spectra were obtained by matching the relative fluxes of the observed spectra to the
photometric B, V and I values.  A range of spectral types are present,
from hot stars with strong Balmer absorption lines to cool stars with
calcium triplet lines in absorption but no Paschen lines.

TiO band emission is prominent in all the spectra.  
In order to see the TiO emission better, the spectra were divided by a
low-order polynomial fit to the continuum.  They were then re-plotted
from 6000--8700\,\AA~ in Figure~\ref{tio_spec}.  The top panel of
Figure~\ref{tio_spec} shows the spectrum of a late M star with the
position of TiO and VO absorption bands marked.  It is clear that for
the objects being studied here, there is TiO emission in most of the
TiO bands in the spectral region shown.  It is not clear that there is
any VO band emission.  The existence of this emission indicates the
presence near the central star of molecular gas with a temperature
less than $\sim$4000\,K and with a temperature profile that increases
toward the observer.  The origin of this temperature structure is
unclear - see \citet{hil12} for a discussion of possibilities.

About half the spectra also show H$\alpha$ in emission (see
Table~\ref{tab1} and Figures~\ref{full_spec} and \ref{tio_spec}).  The
H$\alpha$ emission in J010628.81-715204.8 and J050747.45-684351.2
shows a P Cygni profile corresponding to wind terminal velocities of
approximately 430 and 350 km s$^{-1}$, respectively.
J010628.81-715204.8 also shows the calcium triplet lines in emission.
The H$\alpha$ emission lines in J005504.57-723451.1 and
J052023.97-695423.2 are resolved, and their observed profiles
correspond to the instrumental profile convolved with gaussian
velocity distributions of standard deviation 100 and 180 km s$^{-1}$,
respectively.  The H$\alpha$ emission in J052023.97-695423.2 also has
broad wings extending out to 900 km s$^{-1}$.  We note that H$\alpha$
emission is common in YSOs with disks that are
accreting surrounding gas \citep{app89} but it is also common in
post-AGB stars \citep{van00}.  

In some stars, the [OIII] 5007$\AA$ emission line can be seen.
This is most likely LMC background emission not associated with the object.
Such emission is very hard to remove with a multi-fibre spectrograph
which samples the sky at random points over a two degree field.

\subsection{The stars in the HR-diagram}\label{hrd_sect}

\begin{figure*}
\includegraphics[scale=0.7]{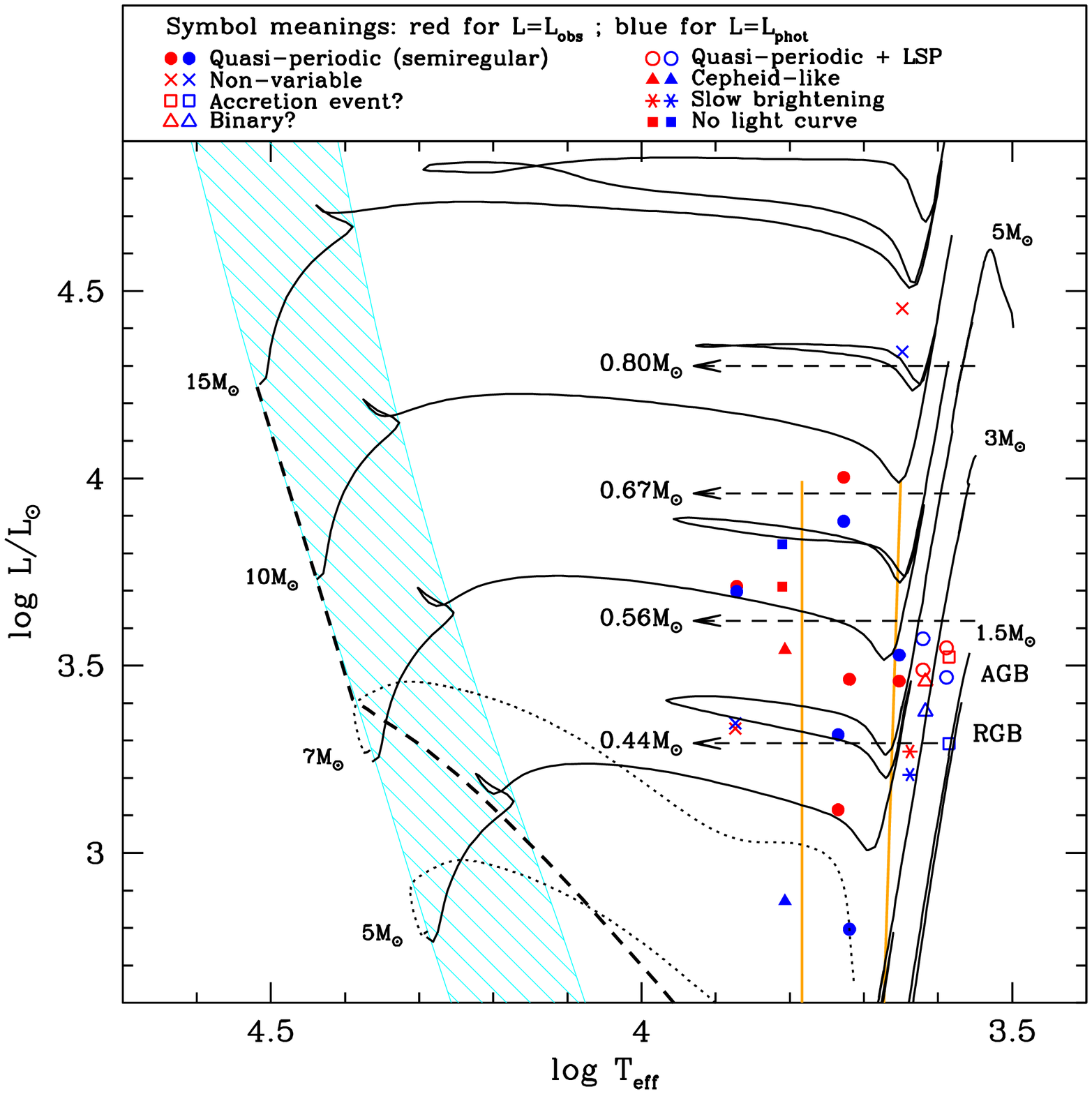}
\caption{The HR-diagram for the stars with TiO bands in emission.
Each star is shown twice, using red symbols for $L_{\rm obs}$ and
cyan symbols for $L_{\rm phot}$.  The symbol shapes depend on the
variability type (see the legend at the top of the figure and 
Section~\ref{var_sec} for details).  Pre-main-sequence
evolutionary tracks from \citet{tog11} are shown as dotted lines
while post-main-sequence tracks from \citet{ber08} and \citet{ber09}
are shown as continuous lines.  The initial stellar masses of the
tracks are shown on the figure.  Only the RGB and AGB phases of 1.5 and
3\,M$_{\odot}$ stars are luminous enough to be seen on the figure.  
Schematic post-AGB and post-RGB  evolutionary tracks
are shown as dashed lines with their masses marked (see text for details). 
The thick dashed line is the birth line
for YSOs from \citet{pal99} with its extension above 7\,M$_{\odot}$
as given by \citet{ber96}.  The cyan hashed region shows the main-sequence
area.  The orange lines delineate the instability strip for 
RV Tauri stars in the LMC from \citet{sos08}.
}
\label{hrd}
\end{figure*}

The stars are shown in the HR-diagram in Figure~\ref{hrd} along with
evolutionary tracks for pre-main-sequence stars, normal
post-main-sequence stars, and post-AGB and post-RGB stars.  Both
$L_{\rm obs}$ and $L_{\rm phot}$ values are shown.  In general, these
two values do not differ greatly compared to the overall spread in
luminosity values, except for the two stars mentioned in
Section~\ref{sed_sec} where the dust is most likely in
a thick edge-on disk or where a second object may be contributing
to $L_{\rm obs}$ thereby making it too large.

For the range of observed luminosities, it is possible to explain the
stars as either pre-main-sequence stars, normal post-main-sequence
stars on helium core burning loops, or post-AGB or post-RGB stars.
However, given that our observed stars all have large amounts of hot
circumstellar material, it is unlikely that they are in the rather
quiet evolutionary stage of helium core burning.  The most likely
stars to have mass loss and an infrared excess in this evolutionary
phase are Cepheid variables.  Yet, observations of these stars in
the LMC \citep{nei10} shows the mid-IR excess flux at 8\,$\mu$m in
these stars is about 10\% of the photospheric continuum whereas in the
present group of objects the mid-IR excess flux at 8\,$\mu$m is
typically about 10 times the photospheric continuum.  We therefore
reject the explanation of these stars as normal post-main-sequence
stars.

The other possible evolutionary states for these objects are post-AGB
or post-RGB star, or pre-main-sequence star\footnote{ In this paper,
we use the term pre-main-sequence star, or PMS star, to describe the
star itself, while we use the term young stellar object, or YSO, to
describe the pre-main-sequence star and its surrounding dust and
gas.  The term post-AGB star or post-RGB star can mean either the
star itself or the star and its surrounding dust and gas. }.  
In both the post-RGB and post-AGB cases, the observed stars
must have undergone a binary interaction in order for them to have
left their respective giant branches.  For RGB stars, which have
luminosities $L < 2500$\,L$_{\odot}$, single-star mass loss is
insufficient to remove the H-rich envelope and produce a post-RGB star
\citep{vw93}.  The only way large amounts of mass loss and evolution
off the RGB can occur is via binary interaction, presumably a common
envelope event that leaves the binary in a tighter orbit than the
original one \citep[e.g.][]{han95}.  For AGB stars, large amounts of mass loss can occur
either by binary interaction or, for single stars, by the very high
mass loss rate ``superwind'' that terminates AGB evolution
\citep[e.g.][]{vw93}.  However, in the Magellanic Clouds, all the single stars
with high mass loss rates, at least for $\log L$/L$_{\odot} < 4.3$,
are carbon stars \citep[e.g.][]{blu06,gro07,gor11}.  Given that the
stars in the present sample are oxygen-rich because they show TiO
molecules in their spectra, we believe that only binary interaction
could explain a post-AGB status.

\begin{figure}
\centering
\includegraphics[width=0.95\columnwidth]{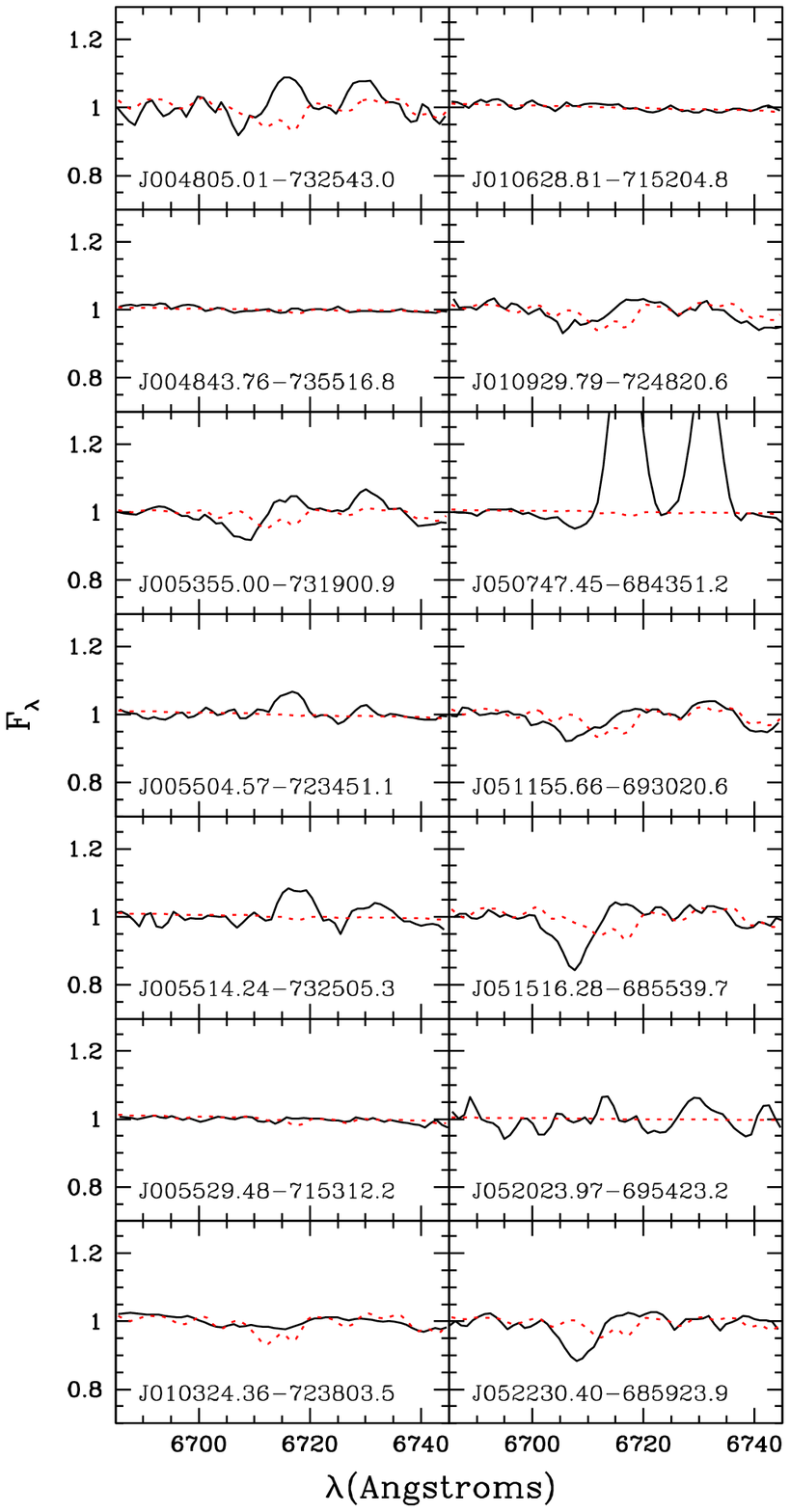}
\caption{A small part of the spectrum of each of the stars covering
the wavelength region of the Li 6708\,\AA~ and the [SII] 6717 and 6731\,\AA~ lines.
Black lines are the observed spectra and red dotted lines are the
best-fit model spectra.}
\label{Li_fig}
\end{figure}

It is not easy to distinguish between post-RGB/AGB binaries and
pre-main-sequence (PMS) stars.  Both are surrounded by large amounts of
circumstellar gas which is accreting (or re-accreting in the case of
post-RGB/AGB stars) onto a circumstellar or circumbinary disk.
However, a distinguishing feature is the gravity of the central star.
We note that, at a given luminosity, the mass of a PMS
star is about 15--20 times that of the corresponding post-RGB/AGB
star.  This leads to a difference of
$\sim$1.3 in $\log g$ between PMS stars and post-RGB/AGB stars.
The errors in our $\log g$ estimates for individual stars
are typically 0.5 \citep{kam13}, although the presence of strong TiO
emission bands in the current set of stars almost certainly increases
the error.  Given the expected $\log g$ difference, 
we can use the derived $\log g$ values in Table~\ref{tab1} to
make a tentative assignment of PMS or post-RGB/AGB
status.  This evolutionary status, PMS or post-RGB/AGB,
is given in Table~\ref{tab1}.  The distinction between post-RGB 
and post-AGB status is based on whether $L_{\rm phot}$ is less than
or greater than, respectively, the RGB tip luminosity $L \approx 2500$\,L$_{\odot}$.  
We estimate that there are 4 post-RGB
stars, 2 post-AGB stars and 6 PMS stars, the latter including the 4
stars hotter than $T_{\rm eff} = 6000$\,K.  For two stars, a status
could not be assigned because the estimated $\log g$ was almost
mid-way between the PMS and post-RGB/AGB values.

Although some of the objects in our sample have been assigned
PMS status on the basis of their gravities, there is a
problem with the existence of optically visible PMS
stars in the part of the HR-diagram where these objects are observed.
PMS stars to the right the ``birth line'' shown in
Figure~\ref{hrd} should not be seen optically
\citep{sta85,pal90,ber96,pal99} since the stars will have evolved to
the birth line or beyond before the optically thick accretion process
has completed.  A possible way around this would be to invoke an
asymmetry in the accretion process that would allow the central star
to be seen through a gap in the surrounding material (for example,
along the pole of the accretion disk).  The birth line shown 
in Figure~\ref{hrd} was computed for Galactic stars with a Galactic Population I abundance.
The lower abundance in the SMC and LMC could make the circumstellar
material more transparent so that PMS stars could be seen
in an earlier phase of evolution.  Indeed, \citet{lam99} and 
\citet{dew05} invoked the lower metallicity of the LMC to explain
the optical visibility of a group of PMS 
candidates in the LMC of O and B spectral type.  We note that our
PMS candidates are much cooler than those of \citet{lam99} 
and \citet{dew05} so that the effect of asymmetry or metallicity
would need to be much more pronounced than for their stars.
The evolutionary status, PMS or post-RGB/AGB,
assigned here should be considered tentative.

At least one of the stars we have classified as a PMS
star, J052023.97-695423.2, has in the past been classified
consistently as a YSO, strongly suggesting a pre-main-sequence status.  
It was observed as part
of the SAGE-SPEC MIR-spectral survey \citep{woo11} and was classified
as a YSO.  It was found to have silicate dust emission, consistent
with the oxygen-rich atmosphere that we have detected.
J052023.97-695423.2 was also classified as a high-probability YSO in
the study of \citet{whi08} which used broad-band near- and mid-IR
colours and luminosities to statistically separate various classes of
objects.

Three of the objects show strong Li lines and several other stars show
marginal detections (Figure~\ref{Li_fig} and Table~\ref{tab1}).
Relatively strong Li lines are found in PMS stars 
only before the Li is destroyed by nuclear
reactions in the stellar interior \citep[e.g.][]{app89}.  If this occurs when the star is
convective to the surface, then the surface Li will be depleted.  The
PMS evolutionary tracks of \citet{ber96} show that
deuterium burning starts early in the evolution of stars with the
luminosities observed here, at $T_{\rm eff} \sim 5000$\,K.  The stars
observed to have strong Li lines (J005355.00-731900.9,
J051516.28-685539.7 and J052230.40-685923.9) have $T_{\rm eff} <
5000$\,K so the presence of the Li lines is consistent with
PMS status.  In Table~\ref{tab1}, one of these stars
(J052230.40-685923.9) is considered a PMS star based on
its $\log g$ value, while the other strong Li sources are a post-RGB star
(J005355.00-731900.9) and a post-AGB star of low luminosity
(J051516.28-685539.7, $L_{\rm phot} = 2943 {\rm L}_{\odot}$).  Strong
Li lines are a feature of AGB evolution at higher masses and
luminosities ($L \ga 20000 {\rm L}_{\odot}$) where hot-bottom burning
occurs \citep{smi90} but we do not expect strong Li lines in 
low-luminosity RGB or AGB stars.  This suggests that J005355.00-731900.9
and J051516.28-685539.7 may also be PMS stars,
contradicting their post-RGB/AGB status based on $\log g$ values.

\section{Variability}\label{var_sec}

\begin{figure*}
\includegraphics[scale=0.70]{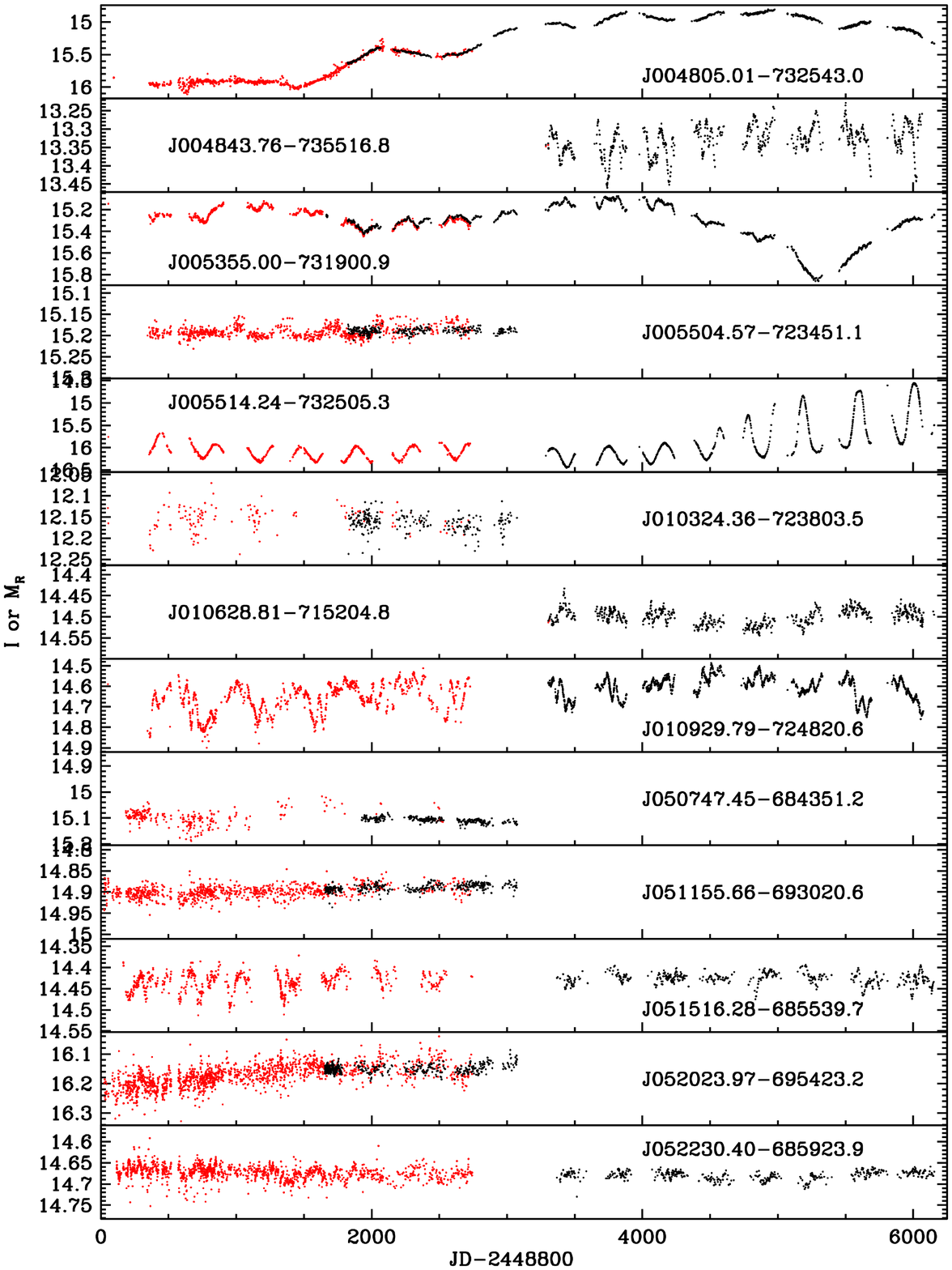}
\caption{Light curves for thirteen of the sources.  The black curves for
dates later than $JD-2448800 > 3250$ are from OGLE III, the
black curves with $1700 < JD-2448800 < 3250$ are from OGLE II
while the red curves with $0 < JD-2448800 < 2800$ are observed MACHO red
magnitudes $M_{\rm R}$ normalized to the OGLE $I$ magnitudes over the interval
$2000 < JD-2448800 < 3250$.}
\label{light_curves}
\end{figure*}

\begin{figure}
\centering
\includegraphics[width=0.9\columnwidth]{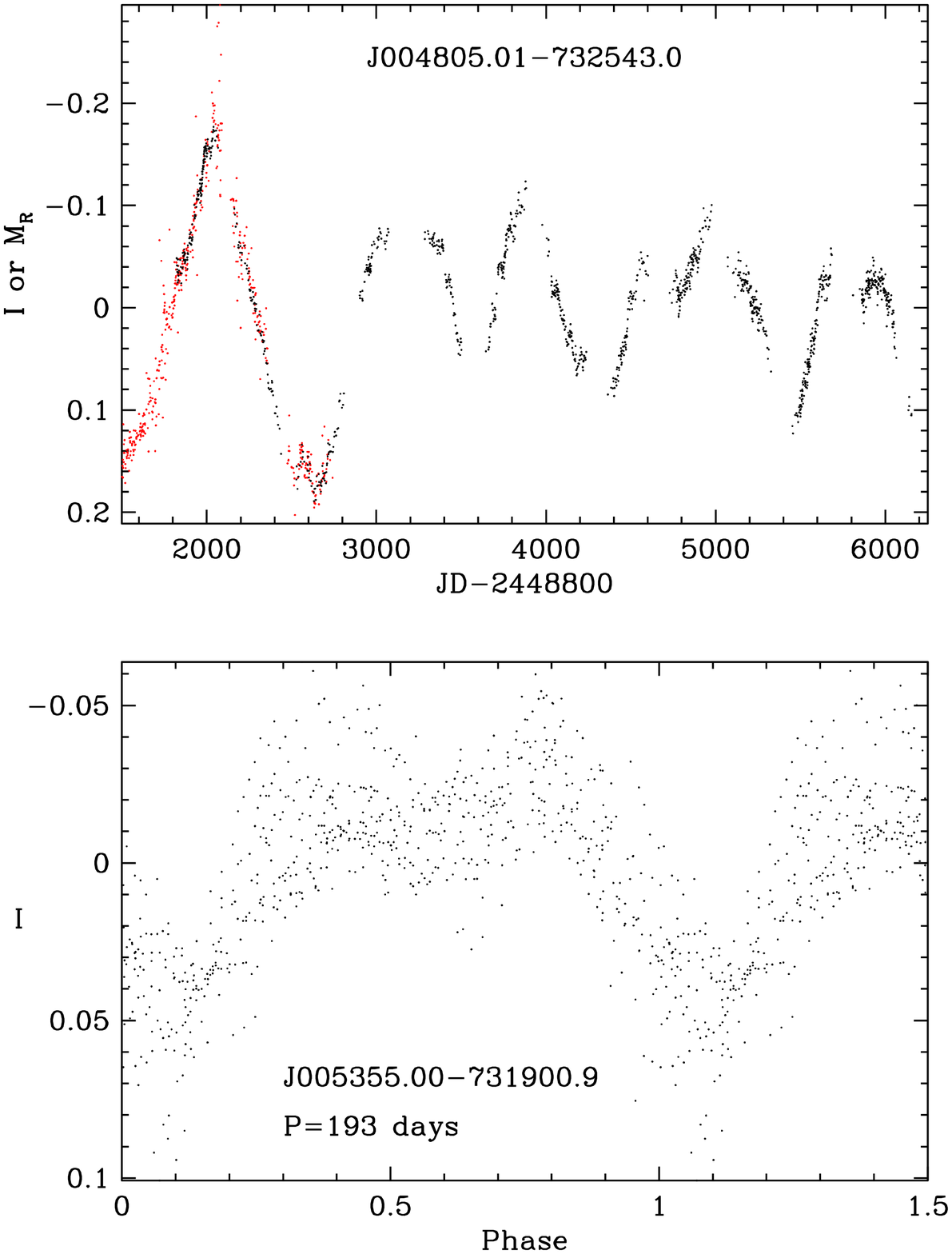}
\caption{The light curves of J004805.01-732543.0 and J005355.00-731900.9
after long term trends have been removed in the intervals $1500 < {\rm JD-}2448800 < 6250$
and $2000 < {\rm JD-}2448800 < 5000$, respectively. The light curve
of J005355.00-731900.9 has been folded with a period of 193 days.  Point
colours are as in Figure~\ref{light_curves}.}
\label{lc_dt}
\end{figure}

Light curves from the MACHO \citep{alc92} and/or the OGLE II and OGLE III
experiments \citep{uda97,szy05, sos09, sos11}  exist for 13 of the 14 stars in our sample, and
they are shown in Figure~\ref{light_curves}.

The stars show a range of types of variability. 
J005504.57-723451.1 and J010324.36-723803.5 are the only stars that
show no detectable variability (the small annual variation that can be seen
in the MACHO observations for J005504.57-723451.1 are an artifact of
the data reduction).  These stars are shown as crosses in Figure~\ref{hrd}.

Most other stars show variability that is a combination of a slow long-term
brightening or fading and variability with periods from 30 days
to more than 3000 days.  The 6 stars J004805.01-732543.0,
J005355.00-731900.9, J005514.24-732505.3, J010628.81-715204.8,
J050747.45-684351.2 and J052023.97-695423.2 all show slow variations
in magnitude over intervals longer than $\sim$1000 days, possibly due to
slow changes in dust obscuration or variability in the rate of
accretion.

J004805.01-732543.0 was fairly constant for the first 1500 days of
observation after which it brightened by about one magnitude
and exhibited an oscillation whose period decreased from
$\sim$1150 days to $\sim$400 days over an interval of 4500 days.  This
oscillation is shown in Figure~\ref{lc_dt} from JD-2448800 =
1500 onwards after the long term trend in magnitude was removed by
fitting a fifth-order polynomial in time.  The oscillation looks RV
Tauri-like (alternating deep and shallow minima) but it is unlikely
that the interior structure of the star associated with this object
could change enough on such a short timescale that a period of
pulsation could change by the observed amount.  It is therefore likely
that this phenomenon is associated with an accretion event in a
surrounding disk, with disk viscosity causing the radius 
of the accreted material and
orbital period to shrink rapidly.  This star is shown as an open
square in Figure~\ref{hrd}.

As well as the slow magnitude changes in J005355.00-731900.9, there is
also an eclipse-like or RV Tauri-like variation with a period between
deep minima of 193 days.  The phased-up light curve in the interval
$2000 < {\rm JD-}2448800 < 5000$ is shown in Figure~\ref{lc_dt} after
the long-term trend was removed by fitting a fifth-order polynomial in
time.  Such a period seems too long for an RV Tauri star:
\citet{alc98} and \citet{sos08} find a maximum period of about 110
days for RV Tauri stars in the LMC and \citet{sos10} find a maximum
period of about 100 days for RV Tauri stars in the SMC (in RV Tauri
stars, the period is taken to be the time between the alternate deep
minima).  We suggest that J005355.00-731900.9 is a binary system with
an orbital period of 193 days.  This star is shown as an open triangle
in Figure~\ref{hrd}.

J005514.24-732505.3 shows a smooth Cepheid-like light curve of increasing
amplitude superimposed on the long-term variation.  The period of
206 days is much longer than that of any Population II
Cepheid in the Magellanic Clouds \citep{alc98,sos08,sos10} but the
star lies close to the instability strip shown in Figure~\ref{hrd} for
these stars.  The rapid increase in amplitude in this star suggests
that it is rapidly evolving through the pulsational instability strip
into a region where growth rates are larger.  One problem with the
interpretation of the 206 day cycle as pulsation is that when the amplitude reaches
$\sim$1 magnitude, the light curve maintains a sinusoidal shape rather than changing to the
saw-tooth shape typical of Cepheids with this amplitude.  It is possible
that the 206 day variation has some other explanation such as binarity.  
This star is shown as a filled triangle in Figure~\ref{hrd}.

The stars J004843.76-735516.8, J010628.81-715204.8,
J010929.79-724820.6, J050747.45-684351.2, J051516.28-685539.7,
J052023.97-695423.2 and J052230.40-685923.9 display
quasi-periodic variability with periods of $\sim$30--160 days.
Their periods are listed in Table~\ref{tab1} when they could be
determined.  Three of the stars are of spectral type K 
where semiregular variability frequently occurs in normal red giant stars but two have
earlier F and G spectral types.  Two of the K stars
(J010929.79-724820.6 and J051516.28-685539.7) also exhibit the long
secondary periods (LSPs) found in roughly one third of variable red
giants \citep{woo99,per03,sosetal07,fra08}.  The 
quasi-periodic variables are shown as filled circles in
Figure~\ref{hrd} unless they have a LSP in which case they are shown
as open circles.

The remaining star with a light curve, J051155.66-693020.6, has slowly
brightening $I$ and MACHO red ($M_{\rm R}$) magnitudes 
with no evidence for other
variability.  It is shown as a star symbol in Figure~\ref{hrd}.  The star
with no light curve (J005529.48-715312.2) is shown as a filled square.

Overall, variability characteristics do not seem to help us
distinguish between PMS stars and post-RGB or post-AGB
stars.  Of the 6 stars that show slow long-term variations in
magnitude, 3 have been classified as post-RGB stars and 3 as
PMS stars based on their estimated $\log g$ values.
The star with Cepheid-like variability is listed as a
PMS star based on $\log g$ whereas Cepheid-like variability is 
known to occur in Population II Cepheids (although at
shorter periods) which are post-AGB or post-RGB stars.  Of the stars
whose quasi-periodic variations look similar to those of red giant
semiregular variables, 3 are classified as PMS stars
and 2 as post-RGB/AGB stars (and two have no classification based on
$\log g$).  It is perhaps not surprising that variability
characteristics are not an unambiguous pointer to the evolutionary
state.  Both PMS stars and post-RGB/AGB stars are expected to be
surrounded by large amounts of circumstellar dust that is clearing so that long-term
slow changes in brightness would be expected in both cases.
Similarly, both groups of stars originate on the Hayashi track 
and cross the instability strip 
so they might be expected to show the
same type of pulsational variability, although the PMS stars
should have shorter periods at a given luminosity because of their higher mass.

\section{Individual object summary}

Given the various results presented above, we provide here a synthesis
of the properties of each object and comment on the possible
evolutionary state.  As a starting point, it should be remembered that
all of these objects have both a strong mid-IR excess and TiO bands
in emission so that they all have dense, warm and dusty circumstellar
material in close proximity to the central star.

J004805.01-732543.0: The most remarkable feature of this star is the
light curve which shows a 1 magnitude brightening starting at $ JD -
2448800 = 1500$ followed by a peak brightness around $ JD - 2448800 =
5000$ and a fading thereafter (Figure~\ref{light_curves}).  During this
phase, there is an oscillation in the light curve with a period that
decreases from 1150 to 400 days.  Such rapid changes in brightness and
period are most likely associated with the outer layers of the star or
the circumstellar environment.  An accretion event in a circumstellar
disk is a plausible explanation.  The star shows H$\alpha$ emission
and this is a possible direct indicator of the existence of a
circumstellar disk.  The estimated effective temperature ($T_{\rm eff}
\sim 3849$\,K) and luminosity (1960--3492\,L$_{\odot}$) put the star
close to the low mass giant branch.  The gravity estimate ($\log g =
0.0$) suggests that the star is a post-RGB star.

J004843.76-735516.8: This star has very strong TiO band emission, no
H$\alpha$ emission, a relatively warm temperature of 5300--5500\,K and
a relatively high luminosity of 7677--10067\,L$_{\odot}$.  The gravity
estimate ($\log g = 0.8$) lies between that expected for a post-AGB or
a PMS star of the given $T_{\rm eff}$ and $L$.  The star
shows a quasi-periodic variability and could be classified as a SRd
variable because of its G0 spectral type.  There is no strong evidence
favouring either post-AGB or PMS status.

J005355.00-731900.9: This is a cool star ($T_{\rm eff} \sim $
4140--4250\,K) whose gravity estimate ($\log g \sim 0.0$) puts it in
the post-RGB/AGB class.  The cool $T_{\rm eff}$, well to the right of
the birth line, also favours a post-RGB/AGB status.  The two luminosity
estimates put it near the RGB tip or just above on the AGB.  It has Li
6708\,\AA\ absorption suggesting it could be a PMS star.  The pointers to
PMS or post-RGB/AGB status for this star are in conflict.  The light
curve has prominent long-term variations of amplitude $\sim$0.6
magnitudes as well as a periodic component.  We suggest that this
star is a binary with a period of 193 days.

J005504.57-723451.1: This is the hottest star in our sample with
($T_{\rm eff} \sim$ 7480--8900\,K), yet it still has prominent TiO
emission.  It has broad H$\alpha$ emission (intrinsic FWHM $\sim$ 235 km
s$^{-1}$) possibly suggesting the presence of a circumstellar disk.
The star does not seem to vary.  The estimated gravity is high ($\log
g \sim 3$) suggesting the star is a PMS star.  The
luminosity of the star ($L \sim$ 2148--2220\,L$_{\odot}$) means that
the stellar mass in this case is $\sim$8\,M$_{\odot}$.

J005514.24-732505.3: This object has two remarkable features.
Firstly, its SED shows a dominant mid-IR peak and a smaller peak in
the optical, indicating a central star that is highly obscured and
seen mainly in scattered light.  Secondly, the star shows a smooth,
periodic, Cepheid-like variation with a period of 206 days whose
amplitude increases from about 0.5 to 1.3 magnitude over an interval
of about 1500 days.  The origin of the 206 day variation is probably
pulsation although some other cause such as binarity can not be
excluded.  The luminosity of the central star is highly uncertain and
the estimates are 743 and 3486\,L$_{\odot}$.  The star is relatively
warm ($T_{\rm eff} \sim$\ 6404--6640\,K) and the gravity estimate of
$\log g = 2.5$ suggests that it is a PMS star.  The luminosity
estimates yield a stellar mass in the range 6--9\,M$_{\odot}$ in this
case.

J005529.48-715312.2: This is a relatively warm star ($T_{\rm eff} \sim$\ 
6457--7690\,K) with a relatively high luminosity ($L \sim$\ 
5141-6668\,L$_{\odot}$).  With the estimated $\log g = 2.1$, it
appears to be a PMS star.  The luminosity
estimate is consistent with a stellar mass of $\sim$12\,M$_{\odot}$ in this
case.  There is no light curve for this star.

J010324.36-723803.5: This is the most luminous of our objects with $L
\sim$\ 21757--28384\,L$_{\odot}$.  It also has a relatively warm
effective temperature $T_{\rm eff} \sim$\ 4447--5500\,K.  The gravity
$\log g = 0.0$ suggests it is a post-AGB star.  This conclusion is
supported by the unusually strong line observed at 6500\AA\ which is
caused by the pair of lines 6496.89\AA\ of BaII and 6498.76\AA\ of
BaI.  The element Ba is produced by the s-process in AGB stars and it
is brought to the stellar surface by the third dredge-up at helium
shell flashes.  The light curve of this star shown no variability.

J010628.81-715204.8: This is the second hottest star in the sample
($T_{\rm eff} \sim$\ 7443--8450\,K) and it is relatively luminous ($L
\sim$\ 4988--5031\,L$_{\odot}$).  The H$\alpha$ line is in emission and
it has a P-Cygni profile suggesting a wind outflow with a velocity
$\sim$430 km s$^{-1}$.  The Ca triplet lines are also in emission.
The estimated gravity ($\log g = 2.5$) indicates that the star is a
PMS star.  In this case, the stellar mass is $\sim$11\,M$_{\odot}$. 
The light curve shows small amplitude variations with a
period of 33.5 days as well as long term variations in mean magnitude.

J010929.79-724820.6: This is a cool star ($T_{\rm eff} \sim$
4000--4173\,K) with a moderate luminosity ($L
\sim$\ 3079--3732\,L$_{\odot}$) that puts it above the RGB tip.  The
TiO emission bands are particularly strong but the mid-IR excess is
relatively weak.  The estimated gravity ($\log g = 0.7$) lie between
that of a post-AGB star and a PMS star.  The light curve is typical of
the semi-regular red giants that show a primary oscillation (in this
case 70--100 days) as well as a long secondary period (in this case
444 days).  We are unable to decide between post-AGB or PMS status for
this star.

J050747.45-684351.2: This a low luminosity object ($L \sim$ 
1304--2068\,L$_{\odot}$) with a warm temperature ($T_{\rm eff} \sim$ 
4850--5430\,K).  The gravity estimate ($\log g = 0.5$) indicates that
it is a post-RGB star.  This object is listed as a YSO candidate by
\citet{whi08}, although not of high probability.  The light curve
shows a slow variation in the mean magnitude and a very small
amplitude oscillation with a period of around 50 days.  The H$\alpha$
line is in emission with a P-Cygni profile consistent with a wind
outflow of $\sim$350 km s$^{-1}$.

J051155.66-693020.6: This is another low luminosity object ($L
\sim$\ 1618--1865\,L$_{\odot}$) with a moderately warm temperature
($T_{\rm eff} \sim$\ 4096--4850\,K) and a gravity estimate ($\log g =
0.0$) indicating that it is a post-RGB star.  The TiO band emission is
relatively weak.  This object is listed as a YSO candidate by
\citet{whi08}, although not of high probability. The light curve shows
a very slow brightening.

J051516.28-685539.7:  This is a cool star  ($T_{\rm eff} \sim 3878$\,K)
of moderate luminosity ($L \sim$\ 2943-3529\,L$_{\odot}$).  The
gravity estimate ($\log g = 0.0$) suggests that it is a 
post-AGB star.  The TiO band emission is relatively weak.  The
star shows variability typical of a red giant semiregular variable with
a primary period of 124 days and a long secondary period of 380 days.
The strong Li 6708\,\AA\ line suggests that the
star may be a PMS star.  This object is listed as a YSO
candidate by \citet{whi08}, although not of high probability.
The post-AGB or PMS status of this object is very uncertain.

J052023.97-695423.2: The luminosity of this object is dominated by the
mid-IR flux, which is still rising at the longest detected wavelength of
24 $\mu$m.  The star appears to be quite warm ($T_{\rm eff} \sim
5244$\,K) and well away from the giant branch. There is a strong and
broad H$\alpha$ emission line of intrinsic FWHM $\sim$ 424 km s$^{-1}$ and
wings extending out to 900 km s$^{-1}$.  Like J005514.24-732505.3, the
two luminosity estimates (626 and 2907 L$_{\odot}$) are quite
different because of the dominance of the mid-IR flux.  The gravity
estimate ($\log g = 2.0$) suggests the star is a PMS star.  The object
has previously been classified as a high-probability YSO by
\citet{whi08}.  As a PMS star, the luminosity estimates suggest a
stellar mass in the range $\sim$7-10\,M$_{\odot}$.
The light curve shows a slowly brightening magnitude
as well as a short period (tens of days) oscillations of low amplitude.

J052230.40-685923.9: This is a cool star ($T_{\rm eff} \sim 4493$\,K)
of moderate luminosity (2876--3375\,L$_{\odot}$) whose gravity
estimate ($\log g = 1.6$) suggests that it is a PMS star.  The strong
Li 6708\,\AA\ line supports the PMS status.  As a PMS star, the
luminosity estimate suggest a stellar mass of $\sim$9\,M$_{\odot}$
although this is very uncertain since such a cool star is on the
steeply-sloped Hayashi track.  It has strong TiO band emission.  The
light curve shows a small amplitude variation with a period of 46
days.  

\section{Summary and conclusions}

We have discovered 14 stars in the Magellanic Clouds that have both a
mid-infrared flux excess and TiO bands in emission in the optical part
of the spectrum.  These features suggest that the stars have dense,
hot dust and gas in their immediate circumstellar environments.  Effective
temperatures have been estimated for the objects from the optical
spectra, reddening estimates have been made and the luminosities of
the central stars derived.  The position of the stars in the
HR-diagram suggest that they are either post-AGB or post-RGB stars of
mass $\sim$0.4--0.8\,M$_{\odot}$ or PMS stars YSOs with
masses of $\sim$7-19\,M$_{\odot}$.  We have tentatively assigned the
stars to one of these two categories based on gravity estimates for
the central stars, although for two of the stars this was not
possible.  We estimate that there are roughly equal numbers of
PMS stars and post-RGB/AGB stars in our sample of TiO emitters.  
Those stars that are PMS stars are in an
evolutionary stage well before the birth line where Galactic PMS stars are
first assumed to become optically visible. Asymmetries in 
circumstellar material or the lower metallicity of the SMC and LMC
may allow PMS stars to be visible in early evolutionary 
stages before the Galactic birth line.  Those that are post-RGB stars
must have formed as a result of binary interaction on the RGB.
Similarly, those that are post-AGB stars must have formed as a
result of binary interaction on the AGB since single AGB stars in the
Magellanic Clouds become carbon stars before they leave the AGB to
become post-AGB stars, yet these stars are all oxygen-rich.  A
circumbinary disk is expected in binary systems that have recently
interacted and such discs can explain the presence of dense warm dusty
circumstellar material in post-RGB and post-AGB stars.

The light curves of  
a majority of the stars show gradual brightening or fading,
consistent with the presence of a changing circumstellar environment.
One of the stars, J004805.01-732543.0,
shows what looks like an accretion event that causes variability with a 
rapidly decreasing period.  The star J005355.00-731900.9
has a light curve suggesting that it is currently an eclipsing binary system
while the star J005514.24-732505.3 has Cepheid-like pulsations 
that are increasing rapidly in amplitude,
suggesting a rapid rate of evolution.  Seven of the stars show
quasi-periodic variability with periods of $\sim$30--160 days.

\section*{Acknowledgments}

We are grateful to the referee, Greg Sloan, whose careful 
comments led to a number of improvements to the paper.
This paper utilizes public domain data obtained by the MACHO Project,
jointly funded by the US Department of Energy through the University
of California, Lawrence Livermore National Laboratory under contract
No. W-7405-Eng-48, by the National Science Foundation through the
Center for Particle Astrophysics of the University of California under
cooperative agreement AST-8809616, and by the Mount Stromlo and Siding
Spring Observatory, part of the Australian National University.
We also acknowledge use of light curve data obtained by the OGLE project as cited
in the text.


\bsp

\label{lastpage}


\begin{thebibliography}{99}

\bibitem[\protect\citeauthoryear{Alcock et al.}{1992}]{alc92} 
Alcock C., et al., 1992, ASPC, 34, 193 

\bibitem[\protect\citeauthoryear{Alcock et al.}{1998}]{alc98} 
Alcock C., et al., 1998, AJ, 115, 1921

\bibitem[\protect\citeauthoryear{Appenzeller 
\& Mundt}{1989}]{app89} Appenzeller I., Mundt R., 1989, A\&ARv, 1, 291 

\bibitem[\protect\citeauthoryear{Bernasconi 
\& Maeder}{1996}]{ber96} Bernasconi P.~A., Maeder A., 1996, A\&A, 307, 829

\bibitem[\protect\citeauthoryear{Bertelli et 
al.}{2008}]{ber08} Bertelli G., Girardi L., Marigo P., Nasi E., 2008, A\&A, 484, 815

\bibitem[\protect\citeauthoryear{Bertelli et 
al.}{2009}]{ber09} Bertelli G., Nasi E., Girardi L., Marigo P., 2009, A\&A, 508, 355

\bibitem[\protect\citeauthoryear{Blum et al.}{2006}]{blu06} 
Blum R.~D., et al., 2006, AJ, 132, 2034

\bibitem[\protect\citeauthoryear{Cardelli, Clayton, \& Mathis}{1989}]{car89} Cardelli J.~A., Clayton G.~C., Mathis J.~S., 1989, ApJ, 345, 245

\bibitem[\protect\citeauthoryear{Castelli 
\& Kurucz}{2004}]{cas04} Castelli F., Kurucz R.~L., 2004, astro, arXiv:astro-ph/0405087 

\bibitem[\protect\citeauthoryear{Cohen et al.}{2004}]{coh04} 
Cohen M., Van Winckel H., Bond H.~E., Gull T.~R., 2004, AJ, 127, 2362

\bibitem[\protect\citeauthoryear{Covey et al.}{2011}]{cov11} 
Covey K.~R., et al., 2011, AJ, 141, 40 

\bibitem[\protect\citeauthoryear{de Ruyter et 
al.}{2006}]{der06} de Ruyter S., van Winckel H., Maas T., Lloyd Evans T., Waters L.~B.~F.~M., Dejonghe H., 2006, A\&A, 448, 641 

\bibitem[\protect\citeauthoryear{de Wit et al.}{2005}]{dew05} de Wit, W.~J., Beaulieu, J.~P., Lamers, H.~J.~G.~L.~M., 
        Coutures, C., Meeus, G., 2005, A\&A, 432, 619

\bibitem[\protect\citeauthoryear{Fraser, Hawley, 
\& Cook}{2008}]{fra08} Fraser O.~J., Hawley S.~L., Cook K.~H., 2008, AJ, 136, 1242

\bibitem[\protect\citeauthoryear{Gordon et al.}{2011}]{gor11} 
Gordon K.~D., et al., 2011, AJ, 142, 102 

\bibitem[\protect\citeauthoryear{Gray 
\& Corbally}{2009}]{gc09} Gray R.~O., Corbally C., J., 2009, Stellar Spectral Classification, Princeton University Press

\bibitem[\protect\citeauthoryear{Groenewegen et al.}{2007}]{gro07} Groenewegen M.~A.~T., et al., 2007, MNRAS, 
376, 313

\bibitem[\protect\citeauthoryear{Han, Podsiadlowski, \& Eggleton}{Han et al.}{1995}]{han95} Han Z., Podsiadlowski P., Eggleton P.~P., 1995a, MNRAS, 272, 800

\bibitem[\protect\citeauthoryear{Hillenbrand et 
al.}{2012}]{hil12} Hillenbrand L.~A., Knapp G.~R., Padgett 
D.~L., Rebull L.~M., McGehee P.~M., 2012, AJ, 143, 37

\bibitem[\protect\citeauthoryear{Kamath et al.}{2013}]{kam13} Kamath, D., Wood, P.~R., \& Van Winckel, H., 2013, to be submitted to MNRAS.

\bibitem[\protect\citeauthoryear{Keller \& Wood}{2006}]{kel06} Keller, S.~C., \& Wood, P.~R., 2006, ApJ, 642, 834 

\bibitem[\protect\citeauthoryear{Lamers, Beaulieu \& de Wit}{Lamers et al.}{1999}]{lam99} Lamers, H.~J.~G.~L.~M., Beaulieu, J.~P., de Wit, W.~J., 1999, A\&A, 341, 827

\bibitem[\protect\citeauthoryear{Meixner et al.}{2006}]{mei06} Meixner, M., Gordon, 
K.~D., Indebetouw, R., et al., 2006, AJ, 132, 2268

\bibitem[\protect\citeauthoryear{Men'shchikov et al.}{2002}]{men02} Men'shchikov, A.~B., Schertl, D., 
Tuthill, P.~G., Weigelt, G., \& Yungelson, L.~R., 2002, A\&A, 393, 867

\bibitem[\protect\citeauthoryear{Munari et 
al.}{2005}]{mun05} Munari, U., Sordo, R., Castelli, F., \& Zwitter, T., 2005, A\&A, 442, 1127

\bibitem[\protect\citeauthoryear{Neilson et al.}{2010}]{nei10} Neilson, H.~R., Ngeow, 
C.-C., Kanbur, S.~M., \& Lester, J.~B., 2010, ApJ, 716, 1136

\bibitem[\protect\citeauthoryear{Palla 
\& Stahler}{1999}]{pal99} Palla, F., \& Stahler, S.~W., 1999, ApJ, 525, 772

\bibitem[\protect\citeauthoryear{Palla 
\& Stahler}{1990}]{pal90} Palla, F., \& Stahler, S.~W., 1990, ApJl, 360, L47

\bibitem[\protect\citeauthoryear{Percy \& Bakos}{2003}]{per03}
{Percy}, J.~R., \& {Bakos}, A.~G. 2003, in The Garrison Festschrift, ed. R.~O.
  {Gray}, C.~J. {Corbally}, \& A.~G.~D. {Philip}, 49

\bibitem[\protect\citeauthoryear{Pickles}{1998}]{pic98} Pickles, A.~J., 1998, PASP, 
110, 863 

\bibitem[\protect\citeauthoryear{Porter \& Rivinius}{2003}]{por03} Porter, J.~M., \& Rivinius, T., 2003, PASP, 115, 1153

\bibitem[\protect\citeauthoryear{Smith et al.}{2004}]{smi04} 
Smith G.~A., et al., 2004, SPIE, 5492, 410 

\bibitem[\protect\citeauthoryear{Smith \& Lambert}{1990}]{smi90} Smith, V.~V., \& Lambert, D.~L., 1990, ApJl, 361, L69

\bibitem[\protect\citeauthoryear{Soszynski et 
al.}{2007}]{sosetal07} Soszynski I., et al., 2007, AcA, 57, 201 

\bibitem[\protect\citeauthoryear{Soszy{\'n}ski et 
al.}{2008}]{sos08} Soszy{\'n}ski I., et al., 2008, AcA, 58, 
293 

\bibitem[\protect\citeauthoryear{Soszy{\~n}ski et 
al.}{2009}]{sos09} Soszy{\~n}ski I., et al., 2009, AcA, 59, 
239 

\bibitem[\protect\citeauthoryear{Soszy{\'n}ski et al.}{2010}]{sos10} Soszy{\'n}ski, 
I., Udalski, A., Szyma{\'n}ski, M.~K., et al., 2010, AcA, 60, 91 

\bibitem[\protect\citeauthoryear{Soszy{\'n}ski et 
al.}{2011}]{sos11} Soszy{\'n}ski I., et al., 2011, AcA, 61, 
217 

\bibitem[\protect\citeauthoryear{Stahler}{1985}]{sta85} Stahler, S.~W., 1985, ApJ, 
293, 207 

\bibitem[\protect\citeauthoryear{Szymanski}{2005}]{szy05} Szymanski, M.~K., 2005, 
AcA, 55, 43

\bibitem[\protect\citeauthoryear{Tognelli, Moroni \& Degl'Innocenti}{2011}]{tog11} Tognelli, P. G., Moroni, P. G., \& Degl'Innocenti, S., 2011, A\&A, 533, A109

\bibitem[\protect\citeauthoryear{Udalski et al.}{1997}]{uda97} Udalski, A., Kubiak, 
M., \& Szymanski, M., 1997, AcA, 47, 319 

\bibitem[\protect\citeauthoryear{van de Steene et al.}{2000}]{van00} van de Steene, 
G.~C., Wood, P.~R., 
\& van Hoof, P.~A.~M., 2000, Asymmetrical Planetary Nebulae II: From Origins to Microstructures, 199, 191 

\bibitem[\protect\citeauthoryear{Van Winckel}{2004}]{van04} 
Van Winckel H., 2004, MmSAI, 75, 766

\bibitem[\protect\citeauthoryear{Whitney et 
al.}{2008}]{whi08} Whitney B.~A., et al., 2008, AJ, 136, 18

\bibitem[\protect\citeauthoryear{Vassiliadis 
\& Wood}{1993}]{vw93} Vassiliadis, E., \& Wood, P.~R., 1993, ApJ, 413, 641

\bibitem[\protect\citeauthoryear{Wood et al.}{1999}]{woo99} 
Wood P.~R., et al., 1999, IAUS, 191, 151 

\bibitem[\protect\citeauthoryear{Woods et al.}{2011}]{woo11} 
Woods P.~M., et al., 2011, MNRAS, 411, 1597 

\bibitem[\protect\citeauthoryear{Wright et al.}{2010}]{wri10} Wright, E.~L., 
Eisenhardt, P.~R.~M., Mainzer, A.~K., et al., 2010, AJ, 140, 1868 

\bibitem[\protect\citeauthoryear{Zaritsky et al.}{2002}]{zar02} Zaritsky, D., Harris, 
J., Thompson, I.~B., Grebel, E.~K., \& Massey, P., 2002, AJ, 123, 855 

\bibitem[\protect\citeauthoryear{Zaritsky et al.}{2004}]{zar04} Zaritsky, D., Harris, 
J., Thompson, I.~B., \& Grebel, E.~K., 2004, AJ, 128, 1606 

\bibitem[\protect\citeauthoryear{Zickgraf et al.}{1989}]{zic89} Zickgraf, F.-J., Wolf, B., Stahl, O., \& Humphreys, R.~M., 1989, A\&A, 220, 206
\end{thebibliography}
\end{document}